\documentclass[10pt,conference]{IEEEtran}
\IEEEoverridecommandlockouts
\usepackage{cite}
\usepackage{array,ragged2e}
\usepackage{amsmath,amssymb,amsfonts}
\usepackage{algorithmic}
\usepackage{graphicx}
\usepackage{textcomp}
\usepackage{xcolor}
\usepackage{booktabs}
\usepackage{multirow}
\usepackage{enumitem}  
\usepackage{verbatim}
\usepackage{subfigure}
\usepackage{caption}
\usepackage{hyperref}
\usepackage{makecell}
\def\BibTeX{{\rm B\kern-.05em{\sc i\kern-.025em b}\kern-.08em
		T\kern-.1667em\lower.7ex\hbox{E}\kern-.125emX}}

\newcolumntype{I}{!{\vrule width 0.5mm}}

\newcommand{\dohang}{\hangindent1em\hangafter1 }
\newcolumntype{R}[1]{>{\RaggedLeft\arraybackslash}p{#1}}
\newcolumntype{L}[1]{>{\RaggedRight\arraybackslash}p{#1}}

\definecolor{mygreen}{RGB}{0, 176, 80}
\definecolor{myorange}{RGB}{250, 142, 0}
\definecolor{myblue}{RGB}{0, 112, 192}
\definecolor{mylightblue}{RGB}{0, 176, 240}
\definecolor{mypink}{RGB}{255, 102, 255}
\definecolor{mybrown}{RGB}{127, 96, 0}
\definecolor{mygrey}{RGB}{109,109,109}
\definecolor{mypurple}{RGB}{112, 48, 160}
\definecolor{mygrey}{RGB}{77, 77, 77}

\begin{document}
	
	\title {Silent Vulnerable Dependency Alert Prediction with Vulnerability Key Aspect Explanation}
	
	\author{\vspace {-5mm}
		\IEEEauthorblockN{Jiamou Sun\IEEEauthorrefmark{1}, Zhenchang Xing\IEEEauthorrefmark{1}\IEEEauthorrefmark{2}, Qinghua Lu\IEEEauthorrefmark{1}, Xiwei Xu\IEEEauthorrefmark{1}, Liming Zhu\IEEEauthorrefmark{1}\IEEEauthorrefmark{3}, Thong Hoang\IEEEauthorrefmark{1}, Dehai Zhao\IEEEauthorrefmark{2}}
		\\
		\IEEEauthorblockA{\IEEEauthorrefmark{1}Data61, Eveleigh, \textit{CSIRO}, Sydney, Australia} 
		\IEEEauthorblockA{\IEEEauthorrefmark{2}Research School of Computer Science, CECS, \textit{Australian National University}, Canberra, Australia} 
		\IEEEauthorblockA{\IEEEauthorrefmark{3}School of Computer Science, \textit{University of New South Wales}, Sydney, Australia} 
		\vspace {-4mm}
		\\ \{Frank.Sun, Qinghua.Lu, Xiwei.Xu, Liming.Zhu, James.Hoang\}@data61.csiro.au \\ \{Zhenchang.Xing, Dehai.Zhao\}@anu.edu.au
		\vspace {-4mm}}
	
	\maketitle
	
	\begin{abstract}
		Due to convenience, open-source software is widely used. 
		For beneficial reasons, open-source maintainers often fix the vulnerabilities silently, exposing their users unaware of the updates to threats. 
		Previous works all focus on black-box binary detection of the silent dependency alerts that suffer from high false-positive rates. 
		Open-source software users need to analyze and explain AI prediction themselves. 
		Explainable AI becomes remarkable as a complementary of black-box AI models, providing details in various forms to explain AI decisions.
		%Noticing importance of Explainable AI, commercial companies including GitHub start to apply Explainable AI in silent dependency alert, proposing CodeQL for maintainers' self-checking.
		Noticing there is still no technique that can discover silent dependency alert on time, in this work, we propose a framework using an encoder-decoder model with a binary detector to provide explainable silent dependency alert prediction. 
		Our model generates 4 types of vulnerability key aspects including vulnerability type, root cause, attack vector, and impact to enhance the trustworthiness and users' acceptance to alert prediction.
		By experiments with several models and inputs, we confirm CodeBERT with both commit messages and code changes achieves the best results.
		Our user study shows that explainable alert predictions can help users find silent dependency alert more easily than black-box predictions.   
		%By far what we know, this is the first study of application of Explainable AI in silent dependency alert prediction, which opens the door of the related domains.
		To the best of our knowledge, this is the first research work on the application of Explainable AI in silent dependency alert prediction, which opens the door of the related domains.
\end{abstract}
	
	\section{Introduction}
\label{sec:intro}
Open-source software (OSS) has been widely used in software supply chain of commercial/non-commercial products.
Meanwhile, open-source software vulnerabilities have been widely spread, causing notable damage in the past \cite{Ponta@2019, Ponta@2018}.
For example, SolarWinds vulnerability of supply chain led to huge damage to the governments and organizations including U.S. and UK governments and Microsoft in 2020 \cite{solarwinds}; 
one year later, Log4Shell, a vulnerability of Apache Log4j that is widely applied in downstream software, affected millions of facilities related to organizations including Amazon, Cloudflare and Tencent, causing apocalyptic disasters \cite{Log4j, Log4j2}.
%\zc{For example, Equifax led to the loss of data over 140 million of U.S. citizens in 2017;
%in the same year, WannaCry and NotPetya also caused severe impacts and money losses to the software users \cite{Xiao@usenix2018, Ponta@2018}. ??Are they caused by supply chain vulnerabilities related to dependency alert? WnnaCry and NotPetya are malware, do they affect supply chain? You should include incidents like log4j which is typical supply chain vulnerability. maybe some examples from https://sysdig.com/blog/software-supply-chain-security/?}
%Therefore, when open-source software vulnerabilities are fixed by the maintainer, it is necessary for users to discover the patches and update the software in time.
Therefore, it is an urgent demand for downstream OSS users to note and patch vulnerabilities in upstream OSS in time. 

\begin{figure}
	%\captionsetup{aboveskip=2pt}
	\centering
	\subfigure[NVD Record with Vulnerability Key Aspects Highlighted]{
		%\captionsetup{skip=0pt}
		\begin{minipage}{0.9\linewidth}
			\includegraphics[width=\linewidth]{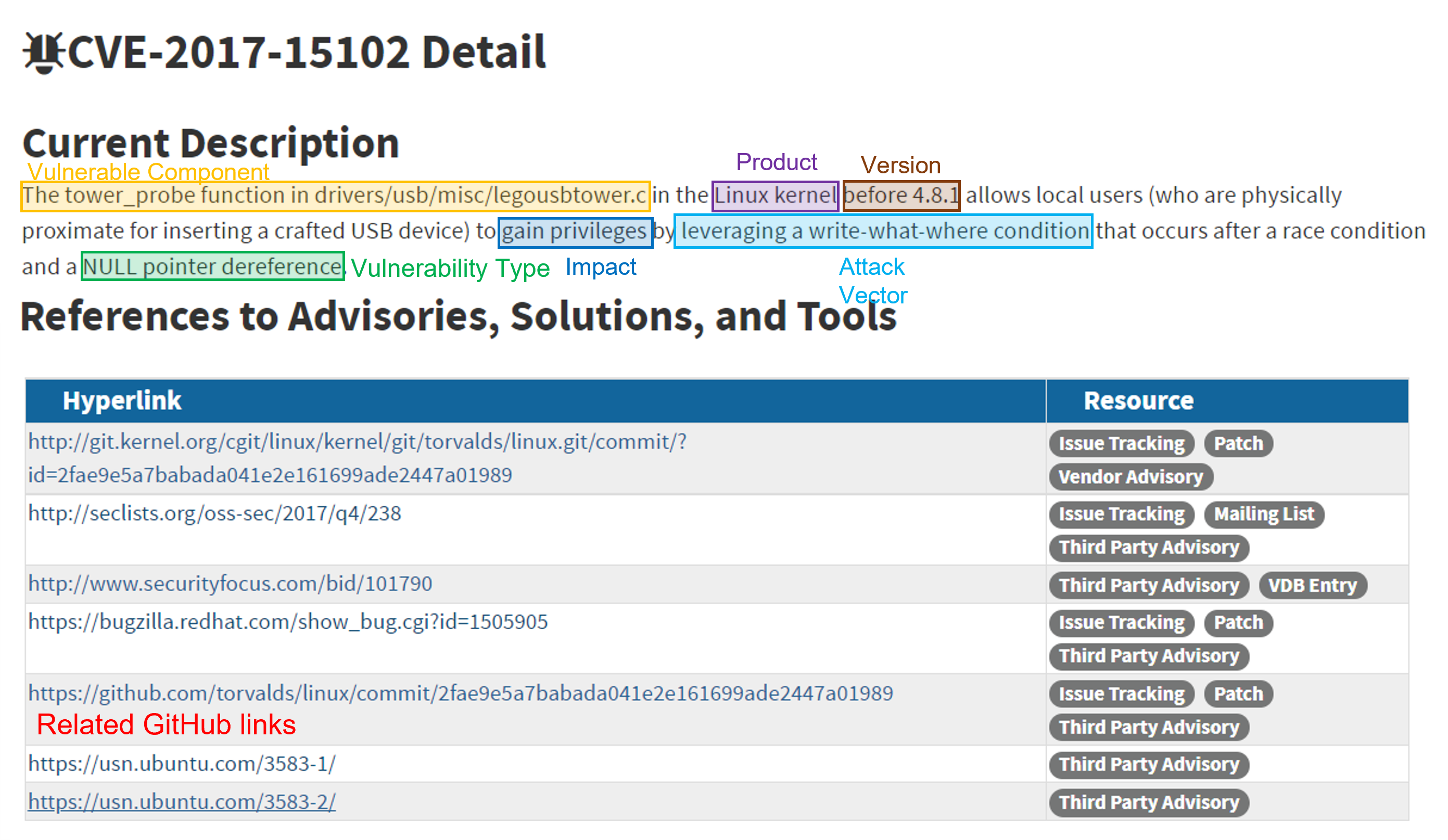}
			\vspace{-2mm}
			\label{fig:nvdexample}
	\end{minipage}}
	\vspace{-1mm}
	\subfigure[Corresponding Silent Vulnerability Patching in Linux on Github]{
		%\captionsetup{skip=0pt}
		\begin{minipage}{0.9\linewidth}
			%\vspace{-1mm}
			\includegraphics[width=\linewidth]{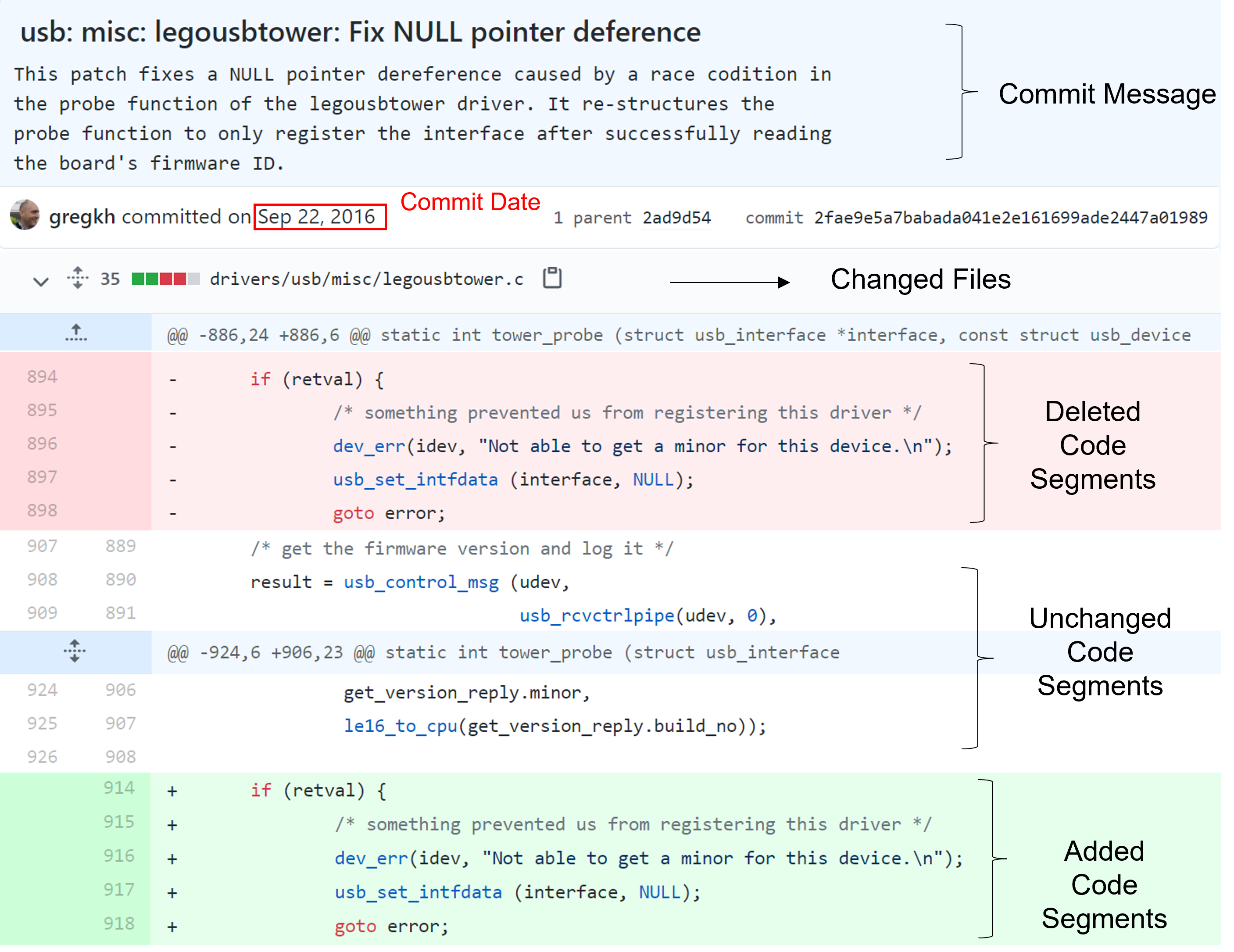}
			\vspace{-2mm}
			\label{fig:github}
	\end{minipage}}
	
	\vspace{-1mm}
	\caption{Examples of NVD Record and Corresponding GitHub Vulnerability Patching Commit}
	\vspace{-6mm}
	\label{fig:examples}
\end{figure}

%One of our industry partners' task is to detect the possible vulnerabilities in their supplychain libraries.
With sheer amounts of OSS libraries, manually finding vulnerability patching updates of dependent software is impossible, so usually, industries rely on public vulnerability advisories to detect vulnerability patching updates.
%Manually find vulnerability patching updates of dependant software in sheer amounts of OSS library is almost impossible, so in general, our industry partner relies on public vulnerability advisories to detect vulnerabilities and patching updates.
Public vulnerability advisory is a dictionary offering unique Common Vulnerabilities and Exposures (CVE) ids to each publicly disclosed vulnerability.
National Vulnerability Databases (NVD) and MITRE are two of the most popular public advisories that accept and review vulnerability submission.
As illustrated in Fig. \ref{fig:nvdexample}, besides CVE id, they also disclose CVE descriptions recording vulnerability-related information, such as vulnerability types, root causes, attack vectors, impacts, and patching information \cite{CVE, NVD}, which provides understandable explanation.
Such information can be used to notify OSS users of the vulnerability characteristics and the releases of software patches.
GitHub Advisory Database \cite{githubadv} is another vulnerability advisory focusing on vulnerabilities of OSS projects in GitHub.
Like NVD, GitHub Advisory Database also records vulnerability characteristics for detailed reviewing.
In addition, GitHub Advisory Database records patching commits with each vulnerability entry.
Through GitHub app Dependabot \cite{Dependabot}, the recorded patching commits can be used for automatic alerts of vulnerable software dependency. 
%However, public vulnerability advisories including NVD and GitHub Advisory Database suffer from the same problem, silent patching.

Although being widely used, public vulnerability advisories including NVD and GitHub Advisory Database suffer from silent patching problems.
Due to unintentional fixing or beneficial considerations, it is not uncommon for OSS maintainers to choose to silently fix vulnerabilities and commit patches without notifying the public \cite{Ruohonen@2018, zhou2021finding}.
For example, Apache's vulnerability handling process suggests developers should not expose vulnerability information related to commits\footnote{https://www.apache.org/security/committers.html}.
As a result, there is a significant delay (on average 77-days in NVD) for vulnerability and patches publishing, or even no publishing \cite{Li@2017,Dong@usenix2019,Sen@2019,Ruohonen@2018}.
Meanwhile, NVD and MITRE only accept vulnerability applications without taking the initiative to discover vulnerabilities. 
Consequently, more than 50\% of the OSS vulnerabilities are silently fixed by the maintainers without recording or with severe recording delays in NVD and MITRE \cite{Wang2020AnES}.
For example, the patch \href{https://github.com/torvalds/linux/commit/2fae9e5a7babada041e2e161699ade2447a01989}{\textcolor{blue}{Commit:7a01989}} in Fig. \ref{fig:github} was released on September 22, 2016, while the corresponding CVE record \href{https://nvd.nist.gov/vuln/detail/CVE-2017-15102}{\textcolor{blue}{NVD:2017-15102}} shown in Fig. \ref{fig:nvdexample} was published on November 15, 2017, which is a year and two months later.
%, which is an example of silent vulnerability fixing on Linux.
%According to the commit message, this vulnerability .
In these cases, downstream OSS users are exposed to high risk due to being unaware of the silently fixed vulnerability.
%, although the software patch has already been released.
Such silent vulnerable software dependencies are highly critical for OSS platforms including GitHub\footnote{https://docs.github.com/en/code-security/repository-security-advisories/about-coordinated-disclosure-of-security-vulnerabilities\#best-practices-for-maintainers}.

To discover vulnerable software dependencies in time, previous works train OSS vulnerability detectors by machine learning to automatically analyze OSS commit messages or code changes \cite{Wang@2020, Zhou@2017,Sabetta@2018}.
These black-box models are capable of automatically recognizing vulnerability patches, then alerting the downstream users.
%Unfortunately, the developers of our industry partner do not trust the classification results, complaining that the lack of the explanation of the classification result forced them to manually inspect the detected code and find the characteristics of the vulnerabilities, such as vulnerable versions, components, vulnerability types, root causes, attack vectors, and vulnerability impacts, in order to make a corresponding vulnerability report. 
Unfortunately, these detectors suffer from unbalanced training data and high false-positive rates, making their prediction results less credible \cite{zolanvari2019machine, Zhou@2017, liu2019deepbalance}.
Because these models only provide binary prediction of patching, they can only show the models' decisions on whether upstream commits are related to vulnerabilities without any explanation.
%, but cannot provide any explanation to the alerts.
Users have to manually check the codes to determine the potential vulnerability characteristics including vulnerability types, root causes, attack vectors, and vulnerability impacts, to have a better understanding of the alerts to confirm or reject them.
This manual analysis can be easily overwhelmed by the sheer amount of dependency alerts. 
%Therefore, our partner still has to manually inspect the predicted-vulnerable dependency updates and analyze vulnerability key aspects.

%Explainable AI (XAI) arises more attention recently, because it not only gives an overview to the model’s decision, but also allows users to view more possibilities of the questions and gain deeper understanding after continually working with the models \cite{bussone2015role}.
Explainable AI (XAI) has gained more attention nowadays. 
It gives an overview of the model’s decision, allowing users to view more possibilities of the questions, and obtain a deeper understanding after continually working with the models \cite{bussone2015role}.
Previous works prove that the interpretation and self-reasoning of XAI benefit users in justifying rationality and working efficiently, enhancing model transparency and human trust, and preventing bias and systems malfunctions \cite{nourani2020don, Mohseni2018ASO}.
%Many works try to figure out the advantages of XAI.
%XX et al. \cite{} claim the interpretation and self-reasoning of XAI could benefit users in justifying rationality and working efficiently when relying on AI decisions. 
%XX et al. \cite{} conduct studies that prove the low model transparency can cause the trust-losing of the users causing self-reliance and disuse of AI capability, and such trust is difficult to be rebuilt once it is lost. 
%XX et al. \cite{} observe the adding of explainable information can significantly improve users’ attitude of the AI models and make the process of working with AI models efficient. 
%XX et al. \cite{} also claims keeping user accountable is critical to the AI development, and users’ comprehension and trust can significantly prevent bias and systems malfunctions. 
Due to its importance, XAI has been widely used in different areas, including social media, e-commerce and data-driven management of human workers \cite{lee2015working, tang2012etrust, tintarev2011designing}. 
%However, such technique has not been used for the OSS dependency alert.

\begin{figure}
	\centering
	\includegraphics[width=0.9\linewidth]{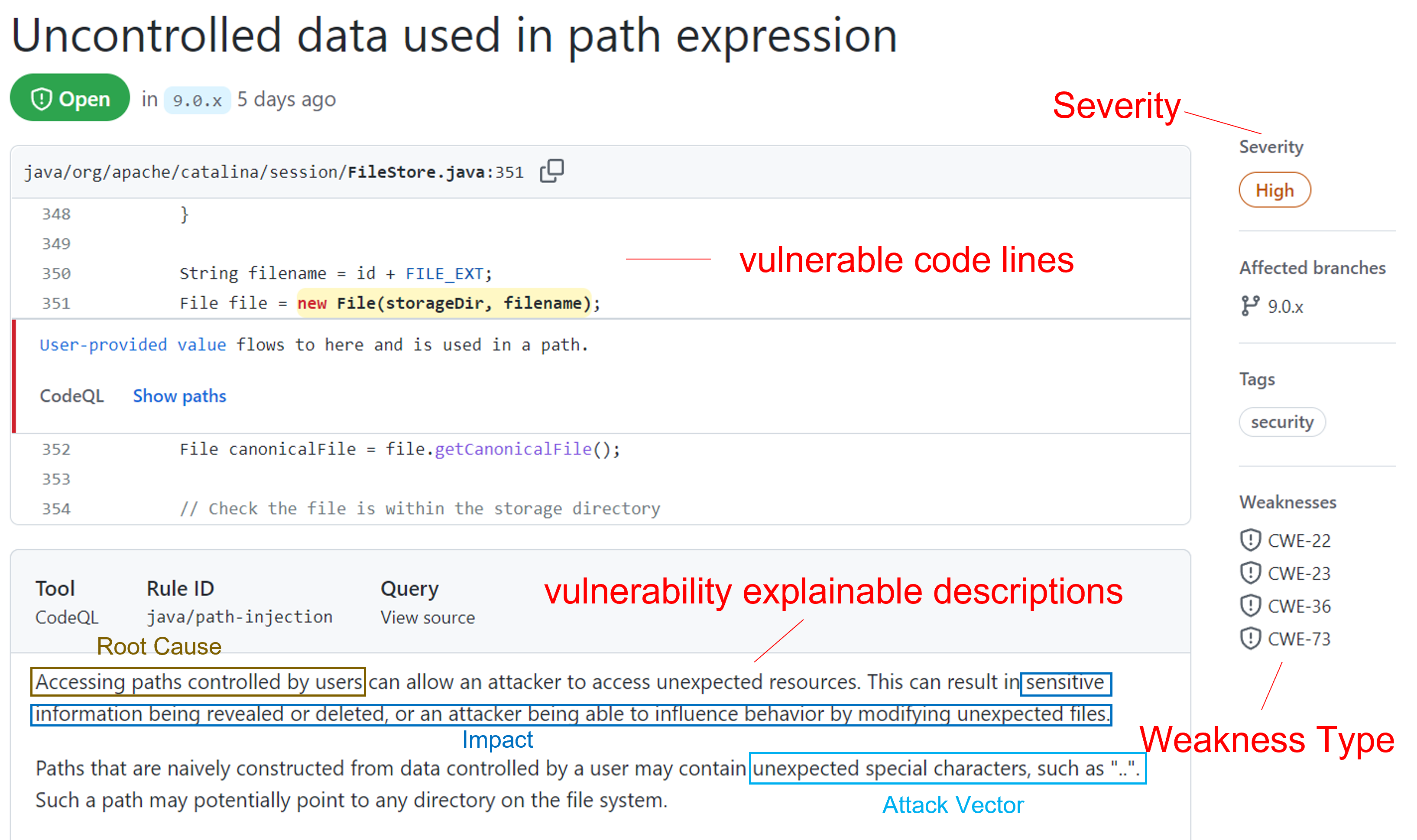}
	\caption{Predicted Alert by CodeQL~\cite{codeql} in Github}
	\label{fig:codeql}
	\vspace{-3mm}
\end{figure}

Noticing the importance of XAI, GitHub develops CodeQL, an explainable vulnerability prediction engine \cite{codeql}. 
CodeQL embedded into the GitHub workflow has the ability to scan code commits, and generate alerts and detailed vulnerability information by both rule-based and machine learning methods. 
Fig. \ref{fig:codeql} shows a screenshot of a predicted alert by CodeQL. 
The alert includes vulnerable code lines, vulnerability severity, weakness types, and detailed explainable descriptions containing vulnerability type, root causes, attack vector, and impact. 
Nevertheless, CodeQL is only for OSS library maintainers to detect the vulnerability risks of their own libraries, but not for the OSS library users to analyze the code updates of dependent libraries.
Unless maintainers decide to publish the alert results, only inner repository people can see the vulnerability alerts and the explanation.
Hence, CodeQL cannot solve the silent dependency patching problem.
Github has another app named Dependabot that is faced by OSS library users for dependency patching alerts.
However, Denpendabot cannot provide any vulnerability information until the vulnerability has been recorded to the GitHub Advisory Database. %which suffers from record delays.
%and can lead to ignorance of OSS library maintainers.

\begin{figure}
	\centering
	\includegraphics[width=0.8\linewidth]{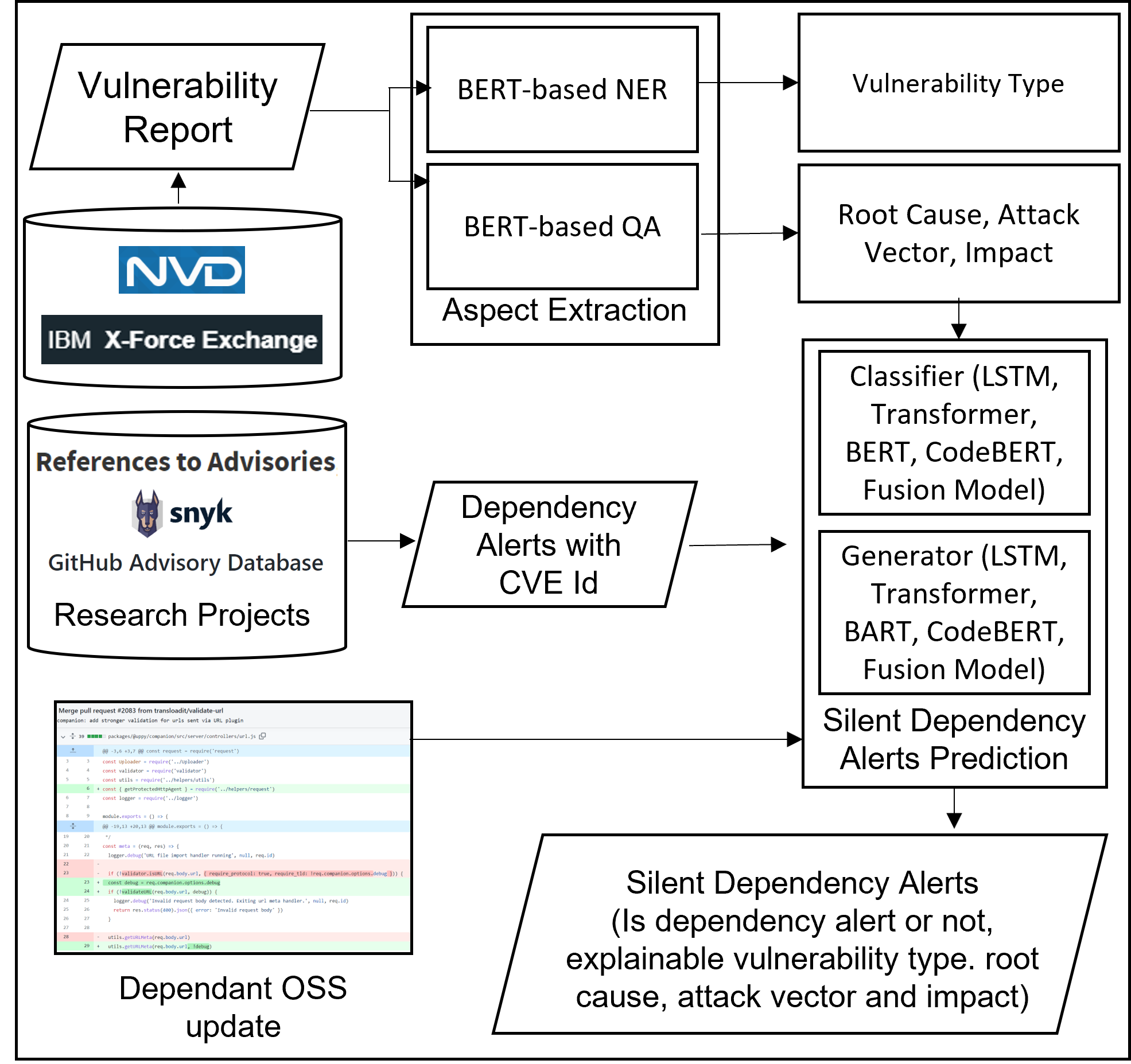}
	%\vspace{-3mm}
	\caption{Overview of Our Frameworks}
	\label{fig:overview}
	\vspace{-5mm}
\end{figure}

In this paper, we propose a framework using encoder-decoder model to predict whether OSS updates in dependent repositories (dependency updates) are vulnerability patches, and generate explainable aspect-level vulnerability information.
Our tool complements the current secure coding practices such as CodeQL and Dependabot.
Fig. \ref{fig:overview} shows an overview of our framework.
{
Previous works produce local explanations by highlighting how inputs contribute to the predictions \cite{9796256, wattanakriengkrai2020predicting, pornprasit2021jitline}.}
However, such method is criticized to be less human-understandable and interpretable \cite{ghassemi2021false, linardatos2020explainable, zhang2020effect, arora2022explain}, so instead, we use human-readable textual information to globally explain our predictions that has been widely used in XAI domains \cite{wiegreffe2021reframing, jimenez2020drug, alonso2020interactive}. 
We follow the NVD templates \cite{CVEtemplate}, using key aspects including vulnerability types, root causes, attack vectors, and impacts as the targeted explainable information, which is coordinated with NVD and GitHub Advisory Database.
We use an extractive method based on BERT Name Entity Recognition (NER) and Question Answering (QA) techniques \cite{sun2021generating, sun2021generating_old} to extract vulnerability key aspects from official descriptions in NVD.
By collecting data from multiple resources, we obtain 8,946 patching commits and their CVEs across 2,017 repositories.
Pairs of patching commits and corresponding CVE key aspects are used to train the prediction and generation model.
%As the vulnerabilities with similar features (i.e., vulnerability type) can have total different characteristics (i.e., other key aspects), the generation of each CVE key aspect is trained separately. 
%for the construction of model training and test data.
%For the classification, we further obtain ??? non-patching commits from the ??? repositories as negative samples.

%We test the effectiveness of classifiers and encoder-decoder models, and use AUC and Rouge scores as metrics, respectively.
{
To evaluate the effectiveness of classifiers and encoder-decoder models, we use AUC and Rouge as our evaluation metrics, respectively. 
Experiments show that CodeBERT \cite{feng2020codebert} with both commit messages and code changes performs the best results compared to various state-of-art baselines, achieving 0.89 AUC and 32.8-45.2 Rouge-1 scores in generation and classification tasks, respectively.
We further conduct a user study offering initial evidence that our explainable dependency alert prediction can help users recognize silent dependency alerts more easily than a black-box binary predictor.}

{
We have made our data avail to the public \footnote{https://github.com/anonymous-dev904/aspect\_generation}.
The repository includes labelled commit patches and 4 types of vulnerability key aspects with the corresponding training code for classifier and generator training, respectively.
In addition, our data includes our experiments results with brief description that are further analysed in the Section \ref{sec:eva}.
}

%In this paper, we make the following contributions:
The main contributions of this paper include:
\begin{itemize}
	\item We provide a large dataset of patching commits and corresponding vulnerability key aspects.
	\item We propose the first approach for predicting silent vulnerable dependency updates with explanations.
	\item We conduct experiments to test the effect of several inputs and models for silent vulnerable dependency prediction.
	\item Our work sheds the light on the XAI to the people's decision on dependency alerts.
	%for vulnerability analysis, demonstrating the importance of vulnerability explanations to the people's decision on dependency alerts. 
\end{itemize}

	\section {Motivating Use Cases}
\label{sec:usecase}

%\zc{I think this section is better placed after the introduction as a motivating scenario.}

In this section, we introduce the functionality of GitHub CodeQL and Dependabot, their limits in XAI and silent dependency alert, and how our explainable silent dependency alert can detect silent patching and give explainable information as a complementary tool for GitHub CodeQL and Dependabot.
%We first give a scenario of vulnerable OSS library and describe how CodeQL works to provide explainable vulnerability detection to OSS library maintainers and its limitation. 
%We then provide details of how Dependabot works to warn the OSS library users after the patch is released on Github Advisory Database and its limitations.
%Finally, we discuss how our explainable silent dependency alert can be used to warn OSS library users while making up the above shortcomings of the two GitHub tools.

\subsection{Vulnerability Detection by CodeQL}
Assume Alice is a maintainer of Apache Tomcat, a popular open-source web server supporting Jakarta Server Pages \cite{Tomcat}. 
In May 2020, Alice activated CodeQL vulnerability detectors to analyze the latest version 9.0.0.x.
%Alice configures the CodeQL workflow to give it the permission to scan code when maintainers merge the pull requests and detect the potential danger with both rule-based and machine learning methods.
She then configured the CodeQL workflow to grant it the code scan permission when maintainers merge the pull requests and detect potential danger with both rule-based and machine learning methods. 
%Alice also used the optional CodeQL machine learning function to detect more vulnerable possibilities than the default rule-based method.
%Alice knew CodeQL can produce false positives, so after scanning, Alice manually checked the results, and found a vulnerability warning shown in Figure \ref{fig:codeql}.
As Alice knew CodeQL might produce false positives, she manually checked the results after scanning and found a vulnerability warning shown in Fig. \ref{fig:codeql}.
From the explainable information including vulnerability types and root cause, She then confirmed this warning was a true positive, which means there was a sensitive leakage risk in the FileStore.java.
Alice committed the patched code \href{https://github.com/apache/tomcat/commit/3aa8f28db7efb311cdd1b6fe15a9cd3b167a2222#}{\textcolor{blue}{Commit:7a2222}} to fix the vulnerability and reported the findings to the NVD on May 6th, 2020. 
However, all the above detection and reporting are only opened within Tomcat.
Before NVD finishes the validation process, the Tomcat users are unaware of the vulnerability and patching, and they have to bear risk themselves.

\subsection{Dependence Alert by Dependabot}
Assume Bill is a downstream user of Tomcat who maintains a problem-spring-web OSS library with over 800 stars on GitHub.
After Tomcat reported the vulnerability, NVD approved and published the vulnerability \href{https://nvd.nist.gov/vuln/detail/CVE-2020-9484}{\textcolor{blue}{CVE-2020-9484}} on May 20th, 2020, which is 14 days later.
Before this time, Bill was unaware of the potential risk of the upstream Tomcat.
Bill continued using GitHub Dependabot to alert the updates of upstream OSS libraries.
On May 12th, 2020, which is 8 days before the vulnerability publication, Dependabot showed there was an updating for Tomcat in the \href{https://github.com/zalando/problem-spring-web/pull/480}{\textcolor{blue}{pull request}}, which actually fixed the vulnerability.
However, there was no explainable information about the vulnerability, hence the update was denied.
On the July 6th, 2020, Dependabot posted \href{https://github.com/zalando/problem-spring-web/pull/484}{\textcolor{blue}{a new pull request}}, which required updating Tomcat library again, was approved by the Tomcat maintainer on July 27th, 2020.
It was 68 and 82 days later after the vulnerability and the patch were published, respectively, hence making a significant delay. 

\subsection{Explainable Silent Dependency Alert}
Assume Bill uses our explainable silent dependency alert prediction tool.
On May 12th, 2020 when Dependabot noticed that there was a commit in Tomcat, Bill captured the commit including the commit message and code changes as inputs to our prediction tool.
The tool predicted that this commit was a vulnerability patch and provided explainable description as below: ``This is patching for \textcolor{mygreen}{directory traversal vulnerability}, the root cause is \textcolor{mybrown}{improper input validation}, attacker can exploit by \textcolor{mylightblue}{sending a specially-crafted request} to \textcolor{myblue}{download arbitrary files on the system}.''
Noticing the potential risk of file leakage, Bill reviewed the committed code.
With help of the explainable description, he found there was no validation process in the FileStore.java before fixing, and the code changes added a validation process, which indicated the prediction was correct.
Found the vulnerable conditions, Bill then updated Tomcat in May, which was much earlier than the real condition.
Note that he possibly updated the vulnerable dependant library even before NVD publishing the vulnerability, and was in a better position to secure the software.
%Hence, they could prove the security environment of the product.
%The scenario above figured out how our tools can be used to protect security of OSS libraries.
%It emphasize the usefulness and importance of our works. 

	\section{Methodology}
\label{sec:method}
This section describes our methodology including training data collection, explainable vulnerability key aspects and extraction, data representatives and diversity, and deep learning models used to generate explanation for dependency alert.

\subsection{Dataset}
%We first introduce our ground truths, the rationality of explainable information selection and our data representatives and diversity.

\subsubsection{Dependency Alert Collection}
We crawled GitHub commits and corresponding CVE ids online.
The vulnerability databases (e.g., NVD) do not require patching commit links as a mandatory requirement when documenting the vulnerabilities.
As such, only a small part of CVE and GitHub patches record their corresponding commit link or CVE id, leading to difficulty for ground truth vulnerability-patch collection.
To overcome the issue, we crawled CVE-GitHub commit links from multiple public databases, including NVD reference lists \cite{NVD}, Snyk databases \cite{Snyk}, Serena et al.'s project \cite{Ponta@2019}, Liu et al.'s project \cite{Liu@icse2020}, and GitHub Advisory Database \cite{githubadv}.

NVD reference lists record supplemental resources including GitHub commits for some vulnerability records (shown in Fig. \ref{fig:nvdexample}).
Snyk is a commercial database aiming for vulnerability alert that contains patching commits of corresponding vulnerabilities.
Serena et al.'s project and Liu et al.'s project focus on vulnerability detection in Java and C dependency alert respectively whose datasets record the traceability of corresponding CVEs and GitHub commits.
%GitHub Advisory Database maintained by GitHub for vulnerability commit detection contains detailed information that has appeared in the specific OSS commits including vulnerable packages, affected versions and commit reference links. 
GitHub Advisory Database is a security vulnerability database containing detailed information, i.e., vulnerable packages, affected versions, and commit reference links, from the world of open-source software (OSS) commits.
These datasets established by different stakeholders for different goals all point out the vulnerability-patch traceability and are synchronized with Mitre CVE entries, which can be used as the vulnerability-patch ground truths in our model training.

\subsubsection{Explainable Vulnerability Key Aspect Collection}
Vulnerability can be described by multi-dimensional information. 
CVE and NVD propose seven human-readable key aspects for vulnerability documentation \cite{CVEtemplate}, which we called explainable vulnerability key aspects. 
As shown in Fig. \ref{fig:nvdexample}, these key aspects are \textbf{name} and \textbf{version} of the vulnerable product, \textbf{vulnerable component} of the product leading to vulnerability, \textbf{vulnerability type} that is commonly identified by the Common Weakness Enumeration (CWE) \cite{cwe} terminology or alike, \textbf{root cause} that describes the program error of the vulnerability, \textbf{attack vector} that describes the attacking signatures, and \textbf{impact} of vulnerability exploit.
%such as cross-site scripting (CWE-79) and XML external entities  (CWE-611), 
%Root cause, attack vector and impact are generally described in free text.

These key aspects, which are widely adopted by different databases and experts \cite{NVD, IBM, Exploitdb, Openwall}, define characteristics of the vulnerability in direct and understandable ways. %and have been used for documentation of CVE entries. 
Such comprehensibility and wide adoption make them efficient references to explain detected vulnerable dependency updates, so we decide to achieve the explanation by using these key aspects.
Because the patching commits already contain product name, version and vulnerable components, in this work we focus on generating vulnerability type, root cause, attack vector and impact from commit messages and code segments.

As illustrated in Fig. \ref{fig:nvdexample}, the explainable key aspects are either formed by named entities (name, version, vulnerable component and vulnerability type) or free-form phrases (root cause, attack vector and impact).
We followed previous works \cite{sun2021generating, sun2021generating_old} that employ either BERT-based NER or QA techniques depending on the aspect characteristics to automatically achieve the extraction of four focused explainable key aspects from vulnerability-related textual descriptions (NVD in this work). 
%NER is straightforwardly designed for named entities, and QA is helpful for the morphological variation of the free-form phrases that can achieve reasonable outcomes of free-form phrase extraction.
{
Previous work \cite{sun2021generating} uses BERT model as its pre-training process under huge textual materials can be helpful on the downstream aspect extraction tasks.
For the entity-based key aspects includes product name, version, vulnerable component and vulnerability type, the work \cite{sun2021generating} uses NER models.
The input of the model is NVD descriptions and the output is the tags of each token.
For the aspects with the morphological variation of the free-form phrases including root cause, attack vector and impact, the work \cite{sun2021generating} uses QA models.
The input of QA model is the concatenation of question including \emph{What is the root cause} with the NVD description, and the output is the probability of start and end spans of the key aspects.
}
As such extractive methods cannot extract vulnerability key aspects directly from code patching commits that usually lack comprehensive description of vulnerability characteristics, the extracted vulnerability key aspects from the NVD descriptions can be used as the ground truth of explainable vulnerability aspects of dependency alerts.

A challenge is that not all NVD descriptions include all four explainable key aspects of our focus.
%Such key aspects are not mandatory in the description, and are hard to be manually determined, especially for the root cause and attack vector, so there are many aspect deficiencies, which affect the model training. 
These key aspects are optional in the description and therefore are hard to be manually determined, especially for the root cause and attack vector. As a result, we have many aspect deficiencies, which affect the model training.
For example, the record \href{https://nvd.nist.gov/vuln/detail/CVE-2019-7263}{\textcolor{blue}{NVD:CVE-2019-7263}} includes only the vulnerable product name and impact information. 
%, and the remaining key aspects are not present.
%As a solution, we used the same BERT-based NER and QA extraction methods to the IBM X-Force dataset \cite{IBM} as compensation.
To overcome this issue, we employ the same BERT-based NER and QA extraction methods for the IBM X-Force dataset \cite{IBM}. 
IBM X-Force maintained by IBM security team that aims for alert system provides detailed and understandable vulnerability information about OSS libraries.
As of August 2021, the IBM X-Force database has collected 145,138 vulnerabilities.
The database is compatible with the CVE documenting standards, using explainable vulnerability key aspects to describe vulnerability.
Unlike NVD, IBM X-Force pays more attention to the completion of documentation details to better support its commercial product, so it has more complete root causes and attack vectors, which can complement the NVD.
As the key aspects of the IBM X-Force database are more general and lack uniqueness compared with NVD descriptions, we prioritize the usage of NVD vulnerability records and key aspects. 
We also take advantage of IBM information when the NVD key aspects are missing.
%we give priority to the usage of NVD vulnerability records and key aspects, and use IBM information when the NVD key aspects are missing.

We obtained the training data from \cite{sun2021generating} and then employed the trained BERT-based NER and QA models to extract the key aspects in the NVD and IBM X-Force descriptions as the ground truth of explainable information. 
%According to \cite{sun2021generating}, the BERT-based NER and QA model only achieve around 0.66-0.90 F1 scores on the vulnerability key aspect extraction from NVD and IBM X-Force.
Both the BERT-based NER and QA models only achieve around 0.66-0.90 F1 scores on the vulnerability key aspects extracting from NVD and IBM X-Force datasets. 
To avoid the impact of false extraction, we further manually validate the extracted key aspects to ensure the correctness of the ground truth data.
{Specifically, two authors manually determine the correctness (binary) of the extracted key aspects of 4,432 vulnerability reports from the NVD and IBM datasets. We use Cohen’s Kappa to measure the agreement. The final Cohen’s Kappa is 0.82, showing perfect agreement as the key aspects are clear in the NVD and IBM descriptions. For those disagreements, the third author will join and solve them together.}

\begin{table}[t]
	\centering
	%\vspace{1mm}
	\caption{Statistics of Dependency Alert Collection}
	\vspace{-1mm}
	\label{tab:sta_dep_alert}
	\setlength{\tabcolsep}{1.5pt}{
	\begin{tabular}{|c|c|c|c|c|c|c|}
		\hline
		
		&C&PHP&Java&Python&JavaScript&Others\\
		\hline
		Rate&38.1\%&16.3\%&12.7\%&4.2\%&4.1\%&24.5\%\\
		\hline
		Number&3,412&1,461&1,136&380&366&2,191\\
		\hline
		
		Avg of Word&\multirow{1}{*}{79.1}&\multirow{1}{*}{19.0}&\multirow{1}{*}{16.0}&\multirow{1}{*}{19.5}&\multirow{1}{*}{36.0}&\multirow{1}{*}{41.6}\\
		\hline
		
		Avg of Added &\multirow{2}{*}{26.0}&\multirow{2}{*}{113.0}&\multirow{2}{*}{128.8}&\multirow{2}{*}{191.8}&\multirow{2}{*}{111.2}&\multirow{2}{*}{17.8}\\
		Code Lines&&&&&&\\
		\hline
		
		Avg of Deleted &\multirow{2}{*}{24.5}&\multirow{2}{*}{58.3}&\multirow{2}{*}{46.6}&\multirow{2}{*}{103.7}&\multirow{2}{*}{37.0}&\multirow{2}{*}{61.8}\\
		Code Lines&&&&&&\\
		\hline

        Avg of Unchanged&\multirow{2}{*}{32.1}&\multirow{2}{*}{65.0}&\multirow{2}{*}{78.6}&\multirow{2}{*}{95.1}&\multirow{2}{*}{76.5}&\multirow{2}{*}{6.7}\\
	    Code Lines&&&&&&\\
		\hline

        Avg of All &\multirow{2}{*}{82.5}&\multirow{2}{*}{236.3}&\multirow{2}{*}{254.0}&\multirow{2}{*}{390.6}&\multirow{2}{*}{224.6}&\multirow{2}{*}{256.2}\\
        Code Lines&&&&&&\\
		\hline
		
	\end{tabular}}
\vspace{-3mm}
\end{table}

\begin{table}[t]
	\centering
	%\vspace{1mm}
	\caption{Statistics of Entry Collection of Each Explainable Vulnerability Key Aspect}
	\vspace{-1mm}
	\label{tab:sta_key_aspect}
	\setlength{\tabcolsep}{1pt}{
	\begin{tabular}{|c|c|c|c|c|}
		\hline
		
		&Vulnerability Type&Root Cause&Attack Vector&Impact\\
		%&&&&\\
		\hline
		
        Number&5,545&5,178&5,752&8,585\\
		\hline

		Avg of Word&\multirow{1}{*}{2.7}&\multirow{1}{*}{6.6}&\multirow{1}{*}{6.9}&\multirow{1}{*}{8.3}\\
		\hline
		
	\end{tabular}}
	\vspace{-5mm}
\end{table}

Table \ref{tab:sta_dep_alert} and Table \ref{tab:sta_key_aspect} show statistics of dependency alert and entry collection of explainable vulnerability key aspects, respectively.
Table \ref{tab:sta_dep_alert} presents the rate of dependency alerts collected, and the average number (Avg) of words of commit messages and added/deleted/unchanged/all code lines by programming language types. 
We totally collect 8,946 pieces of dependency alerts.
Among them, the language types sorted from high to low are C (38.1\%), PHP (16.3\%), Java (12.7\%), Python (4.2\%), JavaScript (4.1\%), and others (24.5\%).
On average, each dependency alert has dozens of commit message words with hundreds of committed code lines, indicating the richness of our training data.

Table \ref{tab:sta_key_aspect} shows the count of entries of each type of explainable key aspect and the average number (Avg) of aspect words.
In total, we have 25,060 pieces of explainable key aspects corresponding to collected dependency alerts from NVD and IBM X-Force.
Compared with other key aspects with on average 6-8 words, the vulnerability type only has around 3 tokens, which is coherent with Sun et al.'s observation \cite{sun2021generating}.

\vspace{-1.5mm}
\subsection{Silent Dependency Alert Prediction with Vulnerability Key Aspect Explanation}

\subsubsection{Silent Dependency Alert Classification}
\label{sec:classification}
%Follow previous work \cite{zhou2021finding} on the same problem, 
We use binary classifiers to achieve silent dependency alert prediction.
Given a patch including commit messages and/or code segments, we predict whether the patch is a vulnerability patch or not by employing five different kinds of classifiers: 
%The list of classifiers is described as follows: 

\textbf{LSTM-based:} 
%Long short-term memory (LSTM) model leveraging cell states with different gates can efficiently adjust weights of contexts for various down-stream tasks, and hence is widely used in different NLP tasks, especially before the Transformer-based models like BERT~\cite{devlin-etal-2019-bert} have been proposed. 
%It has been adopted in silent dependency alert detection in the literatures \cite{zhou2021spi, Zhen@2020}. 
%We choose LSTM as a baseline representing traditional classification model.
Long short-term memory (LSTM) \cite{sak2014long}, composed of cells with different gates, is a type of recurrent neural network. The network has the ability to adjust the weights of contexts for various downstream tasks, hence it is widely used in different natural language processing problems, i.e., handwritten recognition \cite{greff2016lstm}, machine translation \cite{zhou2016deep}, speech recognition \cite{graves2013hybrid}, etc. Moreover, LSTM has been adopted in silent dependency alert detection \cite{zhou2021spi}, \cite{Zhen@2020}, hence we use LSTM as a baseline model in our paper.

\textbf{Transformer-based:} Transformer~\cite{transformer} abandons traditional
structure and uses self-attention mechanism. 
%Each input token is mapped into query (Q), key (K) and value (V) states by weight matrices.
Each input token is mapped into a query state, a key state, and a value state. 
By a normalization function of the query state with the corresponding key state, each value state is assigned to a weight, and the output is calculated by a weighted sum of the value states.
%With success of pre-trained models, Transformer becomes popular in the deep learning domain. 
%We choose Transformer as a baseline of non-pre-trained attention-based model.
%We only use Transformer encoders with a fully connected layer followed by an output layer as the classifier.
As Transformer has become an advanced neural network architecture in the deep learning community, we consider its model as a baseline model. Note that we only employ Transformer encoders with a fully connected layer followed by an output layer as the classifier.

\textbf{BERT-based:} BERT is a Transformer-based pre-trained language model built on large textual corpora with masked word prediction and next sentence prediction as the learning objectives.
%Such pre-trained language models have shown significant improvements of deep learning models in NLP, including BERT NER and QA.
%and led the wave of pre-training in deep learning. 
%Because of pre-training under huge amount of data, it has acquired SOTA in many NLP tasks including classification, named-entity recognition etc. 
%We choose BERT as a representative pre-trained language model for our prediction task.
These pre-trained language models, such as BERT Name Entity Recognition (NER) and Question Answering (QA), have shown significant improvements in NLP problems \cite{sun2021generating}, \cite{sun2021generating_old}. For this reason, we choose BERT as a representative pre-trained language model for our prediction task. 
We connect a fully connected layer to the BERT's [CLS] head followed by an output layer for classification.

\textbf{CodeBERT-based:} CodeBERT has the same structure as the BERT, and it introduces code-specific pre-training and has been shown to achieve the SOTA in many code-related downstream tasks~\cite{feng2020codebert,zhou2021finding,roziere2020unsupervised,NEURIPS2021}. 
%CodeBert is pre-trained with both code and corresponding code comments.
It has been used in Zhou et al.’s work \cite{zhou2021finding} for silent dependency alert detection, in which a fully connected layer is added to the BERT's [CLS] head followed by an output layer for classification.
We reuse its model implementation as a representative baseline of code-specific pre-trained models.

\begin{figure}
	\centering
	\includegraphics[width=0.8\linewidth]{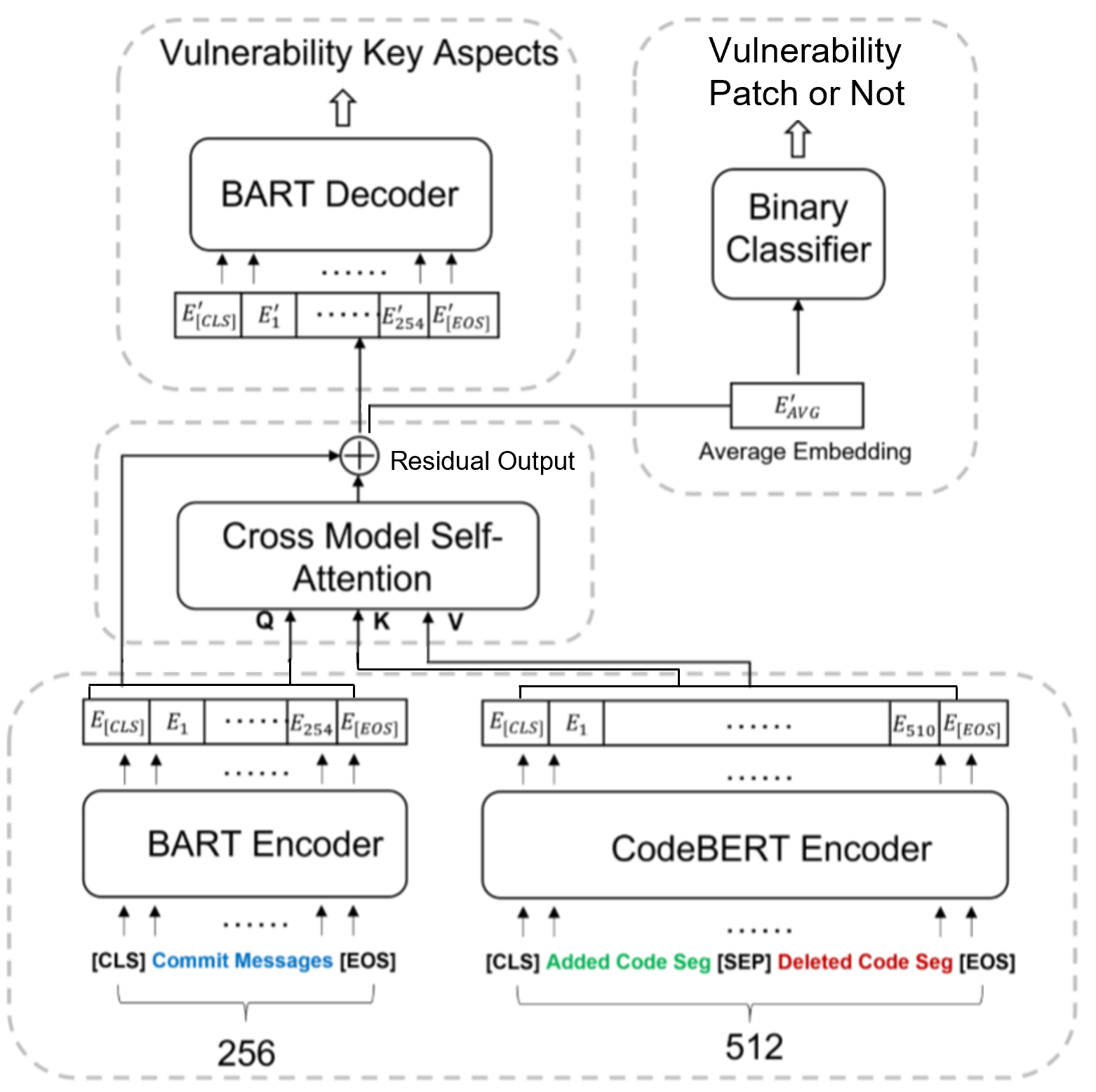}
	%\vspace{-3mm}
	\caption{Our Model of the Transformer based Encoder-Decoder with Cross-Model Self-Attention Layer} %\zc{1) ``Vulnerability Patch or Not'' as the classifier output? 2) Need to explain the model details in the discussion: What is Q, K, V? Why two arrows K and V from CodeBert encoder, but just one arrow Q from BART encoder? Is the whole vector fed into cross model self attention or just some parts or elements of the vector? If it is the whole vector, use a bracket to indicate so. If it is a specific vector element like E[cls], then use a separate line from that element. Use separate lines if K and V are different things. What is the line from E[cls] of BART to $\oplus$? Is $\oplus$ concatenation or element addition? Is E'[avg] computed from E'[cls]...E'[eos]? If so, the line should from E'[cls]...E'[eos], not from the middle of a line. In general, use seprate lines to clearly indicate the source and the target. Do not use a line split into multiple target, or a line from a line.}}
	\label{fig:network}
	\vspace{-6mm}
\end{figure}

\textbf{BERT-CodeBERT-based}: Finally, we develop a model structure combining the pre-trained BERT with the pre-trained CodeBERT, whose structure is shown in Fig. \ref{fig:network}.
GitHub commit contains both text and codes, hence using models with pure NLP and code-specific pre-training together may produce better results. 
%As shown in Figure \ref{fig:network}, 
We add a cross-model self-attention layer to make the two models compatible with each other for our explainable key aspects prediction task, which is detailed in Section \ref{sec:generation}.
Note that we only need the BERT encoder (not the BART encoder-decoder explained below) for the vulnerability patch classification task.

The classifier input contains commit message and code segments (see Fig. \ref{fig:github}), we then have five input variations.
First, we can use only the commit message that may contain textual explanations, however we will lose code information.
Second, we can use only the code changes that contain vulnerability fixing and lose commit message information.
Third, we can use all of the code information including code changes and unchanged code.
Forth, we can combine commit message and code changes together to consider more information.
Fifth, we can use commit message and all of code information as the input.
For all the classifiers, we use the Softmax function \cite{bridle1990probabilistic} for classification.
We fine-tuned all BERT-based models and use Word2Vec vectors~\cite{mikolov2013distributed} as the initial word vectors of non-pre-trained Transformer and LSTM.

\subsubsection{Explainable Dependency Alert Generation}
\label{sec:generation}
We use an abstractive encoder-decoder model to achieve the generation of four explainable key aspects given a patch. 
Encoder-decoder model is widely used in text generation tasks.
The encoder can encode the input to the latent vectors, and the decoder can decode latent vectors into human understandable tokens.
To be consistent with the classification task, we also experiment with five different encoder-decoder models.
\textbf{LSTM-based:} we use two LSTM models as the encoder and decoder to test the capability of traditional deep learning models.
The input of LSTM decoder is the concatenation of attention-based pre-generated outputs and latent variations of the encoder. 
\textbf{Transformer-based:} we use the Transformer encoder-decoder model~\cite{transformer} to test the capability of a non-pre-trained Transformer.
\textbf{BART-based:} 
%BART model \cite{lewis2019bart} combined with pre-trained BERT encoder with pre-trained Transformer decoder has achieved the SOTA results for many generation tasks.
we use the BART model combined with a pre-trained BERT encoder and a pre-trained Transformer decoder to test the capability of the pre-trained Transformer encoder-decoder model.
\textbf{CodeBERT-based:} we use a pre-trained CodeBERT encoder with a non-pre-trained Transformer decoder model to test the capability of the pre-trained code-specific encoder-decoder model.
\textbf{BART-CodeBERT-based}: Similar to the BERT-CodeBERT-based classifier, we combine the pre-trained BART model with pre-trained CodeBERT model 
%whose structure is shown in Figure \ref{fig:network} 
to test the capability of the model containing both natural language and code information.
%As mentioned in Section \ref{sec:classification}, using natural language and code information together may achieve better results.
Unfortunately, one drawback of the pre-trained Transformer model is that the dimensions of its input and parameters are fixed, so we cannot simply concatenate the two encoders' outputs, which is incompatible with the dimensions of the pre-trained decoder's input.
To solve the problem, we use a cross-model self-attention layer, using query states (Q) of BART encoder with key (K) and value states (V) of CodeBERT encoder to calculate the importance of each CodeBERT output token to the BART encoder's output tokens.
The input of BART the decoder is an output of residual structure \cite{he2016deep} between the BART encoder and the cross-model self-attention layer.

The input of the generation is the same as the classifier, which is described in Section \ref{sec:classification}.
The output is the explainable vulnerability key aspects of the vulnerable code patch.
As the vulnerabilities with the same vulnerability type can have totally different root causes or impacts, we train the encoder-decoder model for four kinds of key aspects separately to prevent mutual interference of different aspects.

{Note that the data used to train the classification and the generation models does not depend on the progression of commit time, that is, we do not train the models to specifically predict the condition of future patches.
This is because the features of vulnerable code and patches remain stable and do not change dramatically over time.
For example, in the \href{https://github.com/ushahidi/Ushahidi_Web/commit/4764792d628d40d061f910102135b33ae2123c81}{\textcolor{blue}{Commit:2123c81}} in 2012, the code is vulnerable to SQL injection due to lacking escape of special characters (escape() in line 124).
The similar issue happened in two different projects in  \href{https://github.com/centreon/centreon/commit/434e291eebcd8f56771ac96b37831634fa52b6a8}{\textcolor{blue}{Commit:a52b6a8}} and the \href{https://github.com/phpmyadmin/phpmyadmin/commit/c86acbf3ed49f69cf38b31879886dd5eb86b6983}{\textcolor{blue}{Commit:86b6983}} in 2014 and 2020, respectively (escape () in line 156 and escapeString () in line 3070 of the two commits, respectively).
They all lack the necessary escaping steps before executing SQLs, and the time spans over nine years.
According to the statistics from CVE Details \cite{CVEDetail}, there are 10,871 SQL injection CVEs with the corresponding patches that continue emerging in the past 24 years. 
Therefore, it is reasonable to use the patches from any periods to train the models, so the models will be able to make predictions and explanations for the patches before or after the periods used for model training.}

	\section{Evaluation}
\label{sec:eva}
This section reports the effectiveness of our classifier and generator. 
We present the following three questions:

\begin{itemize}[leftmargin=*]
	\item RQ1: How is the accuracy of silent dependency alert classification?
	\item RQ2: How is the accuracy of explainable vulnerability key aspect generation?
	\item RQ3: How useful is our explainable silent dependency alert prediction compared with only binary patch classification?
\end{itemize}

{In RQ1 and RQ2, we conduct an ablation study to understand the impact of input and model variations to the alert classification and explainable information generation.
The input variants are to determine the minimum information requirement among five information aspects available in the commits for the proposed technique to be effective.
As the transformer models have limited length in the input context window, it is important to validate the necessary contents that should be involved and the compromise that can make when the length of content window is changed.
Similarly, the pre-trained large language models like BERT and CodeBERT can be computational intensive and resources demanding.
The model variants help users to have a full view about influence of different computing resources to the model effectiveness.
In brief, our ablation studies provide a clear picture of adopting our method under different situations. 
}

For all models in the detection and generation, including LSTM, BERT/BART, CodeBERT, BERT/BART-CodeBERT, and Transformer, the input embedding vector is set to 256 dimensions, and the hidden layer dimensions are 768.
The learning rate of LSTM is set to 1e-4, while the rates of other models are 1e-5. 
%For the LSTM model in the detection and generation, the input embedding vector is set to 256 dimensions, the hidden layer dimensions are 768, which is same as Transformer model. 
%The learning rate of LSTM is set to 1e-4. 
All evaluations are done on an Intel Core i9-9900K CPU and a Nvidia GeForce 1080Ti GPU.

\subsection{RQ1: Accuracy of Silent Dependency Alert Classification}

We validate the effectiveness of models (Zhen et al.'s LSTM \cite{Zhen@2020}, Transformer, BERT, Zhou et al.'s CodeBERT \cite{zhou2021finding}, BERT-CodeBERT), and the combinations of inputs (using only commit message, using only added and deleted code segments, using all code segments, using commit message and added and deleted code segments, and using commit message and all code segments) mentioned in Section \ref{sec:classification}.
For the input with only commit messages, only added and deleted code segments, and only all code segments, there are only commit messages or code changes, hence the empty part will be left blank.

\subsection{Dataset}

The dataset includes 1,646,245 commits collected from 2,017 GitHub repositories. 
Among them, 8,946 commits are vulnerable patching commits, and 1,637,299 commits are non-vulnerable patching commits. 
We crawl the commits from the same repositories to simulate the real situation of imbalanced vulnerable versus non-vulnerable patching commits in vulnerable dependency update prediction.
{
We perform 5-fold cross-validation, using same amount of vulnerable data with random-selected down-sampling non-vulnerable data from the same repositories as the training data. 
We use vulnerable commits from the rest repositories with all the non-vulnerable commits of the same repositories as the tested data. 
}
As the testing data is extremely imbalance, 
we use AUC (Area Under The Curve) to measure the classification effect \cite{zhou2021finding}. 
%we follow Zhou et al.’s experiment setting \cite{zhou2021finding}, using AUC (Area Under The Curve) to measure the classification effect.
%\zc{Explain how to compute AUC and why it is better than accuracy in imbalanced data ...}

\begin{comment}
    AUC is calculated from ROC (Receiver Operating Characteristics) \cite{bradley1997use}.
ROC is a curve reflecting the relation between true positive rate and false positive rate of the model.
AUC is the area under ROC curve in the axis that ranges from 0.0 to 1.0.
The larger the AUC score is, the better the model is.
Because AUC considers both true and false condition of positive prediction, it is suitable to measure effectiveness of imbalanced data training.
\end{comment}

\subsubsection{Effectiveness of Different Models}
\label{sec:class_model}

	\begin{table}
	\centering
	%\vspace{1mm}
	\caption{{Results of Silent Dependency Alert Classification by Different Models}}
	\vspace{-1mm}
	\label{tab:result_class_model}
	\setlength{\tabcolsep}{2.5pt}{
	\begin{tabular}{|c|c|c|c|c|c|}
		\hline
		
		&CodeBERT&BERT&BERT-CodeBert&Transformer&LSTM\\
		\hline
		
		AUC &\textbf{0.89}&0.79&0.80&0.66&0.71\\
		\hline
	\end{tabular}}
	\vspace{-4mm}
\end{table}

%Through Section \ref{sec:class_input}, the effectiveness of using commit message and all code segments is the best, so we use this kind of input in the RQ2.

%\zc{How does our model compare with Jiayuan's ASE? Need to methion. In addition to the overall comparable results,it would be better if we could provide some deeper analysis (e.g., language-specific analysis) ... }

{
We use the most comprehensive input (i.e., commit message and all code segments) to investigate the model's effectiveness.
Table \ref{tab:result_class_model} shows the results of classification. 
CodeBERT has the best result (0.89 AUC), followed by BERT-CodeBert (0.80 AUC) and BERT (0.79 AUC).
They both have much better results than non-pre-trained models (Transformer and LSTM), demonstrating that pre-trained models have significant advantages in silent dependency alert classification.
The result of CodeBERT is slightly better than BERT, showing the usefulness of the code-considered pre-training process for code update prediction.
Surprisingly, the BERT-CodeBert model has worse performance than pure CodeBERT.
Our analysis suggests this is because the pre-trained BERT and CodeBERT models are not fully compatible with each other.
During different pre-training processes, the data distribution of the latent parameters may differ greatly, so the two models can output different embeddings, which makes it difficult for the BERT classification head (multi-layer DNN with dropout) to capture the meaningful features.
Another reason is the lack of code data and the deep layers with complex parameters affect negatively the BERT-CodeBert model, especially in our task, whose training set combines different kinds of vulnerabilities and programming languages.
Under such situation, using CodeBERT alone is better than training a complex BERT-CodeBert model.
Compared with Transformer (0.66 AUC), the non-pre-trained LSTM-based model is better due to less layers and parameters required (0.71 AUC).
%With self-attention technique, the Transformer model can better mode the importance of commit messages to the code segments and vice versa.
}

\begin{comment}
	Surprisingly, the BERT-CodeBert model (i.e., the fusion of separate natural language and code embeddings) has the worst performance, with only 0.57 AUC.
	Our analysis suggests this is because the pre-trained BERT and CodeBERT models are not compatible with each other.
	During different pre-training processes, the data distribution of the latent parameters may differ greatly, so the two models can output different embeddings, which makes it difficult for the BERT classification head (multi-layer DNN with dropout) to capture the meaningful features.
	Another reason is the lack of code data and the deep layers with complex parameters affect negatively the BERT-CodeBert model, especially in our task, whose training set combines different kinds of vulnerabilities and programming languages.
	Under such situation, using CodeBERT or BERT alone is better than training a complex BERT-CodeBert model.
	Note that pre-training of CodeBert involves both code and corresponding code comments which makes it work well for the mixed input of commit messages and code segments.
	%Our results show that the text and code information can be more coherently fused in one model which works much more effectively than the late cross-model fusion of separate text and code embeddings.
	
	%, so the Transformer can perform better that traditional LSTM though they are both untrained. 
\end{comment}

\vspace{1mm}
\noindent\fbox{\begin{minipage}{8.6cm} {Pre-trained classifiers are much better than non-pre-trained classifiers. 
CodeBert turns out to be more effective than BERT and fusing the two embeddings in one model.}
%The cross-model fusion of separate commit-message and code-segment turns out to be ineffective than fusing the two embeddings in one model. 
\end{minipage}}\\

\subsubsection{Effectiveness of Different Input Combinations}
\label{sec:class_input}
\begin{table*}
	\centering
	%\vspace{1mm}
	\caption{{Results of Silent Dependency Alert Classification with Different Types of Inputs}}
	%\zc{1) Why not experiment other classifiers? Maybe present model results first then input results? 2) In addition to the current overall results, would it be possible to show the results by each language for Table 1 and Table 2? if the results by each language are intuitive and consistent or reveal some interesting language-specific points. The problem for this RQ is that it is essentially the same as Jiayuang's ASE2021. In addition to the overal comparable results, we need to show some deeper results and analysis. }
	\vspace{-1mm}
	\label{tab:result_class_input}
	\setlength{\tabcolsep}{2.5pt}{
	\begin{tabular}{|c|c|c|c|c|c|}
		\hline
		
		&\multirow{1}{*}{Commit Message}&Added \& Deleted Codes&All Codes&Commit Message \& Added \& Deleted Codes&Commit Message \& All Codes\\
		\hline
		
		AUC &0.82&0.79&0.78&0.84&\textbf{0.89}\\
		\hline
	\end{tabular}}
	\vspace{-2mm}
\end{table*}

%We choose \zc{CodeBERT as the evaluation target in this RQ as CodeBERT is pre-trained under natural language texts and code contents, which can work well under different input combinations ??do not feel this is a good justification to use only CodeBert? It would be important to see the results in Table 1 are consistent for different models. Maybe move 3.2.2 before this experiment. Then, we can say CodeBert is the best model so we use only CodeBert for different inputs.}.
{
From Section \ref{sec:class_model}, CodeBert performs the best under the most comprehensive input, hence we use CodeBert for evaluation.
Table \ref{tab:result_class_input} summarizes the results of the CodeBERT under various input types. 
Overall, the input results with both commit messages and code segments are better than the results without messages or code segments (0.84-0.89 AUC versus 0.78-0.82 AUC), indicating that both commit messages and code segments are useful for the silent dependency alert detection.}

Such results are consistent with our observations.
Some vulnerability patching will directly show vulnerability messages. 
%or \zc{even CVE id in the commit message ??should not mention this? it contradicts the silent prediction setting if a commit already has the CVE id}.
For example, in the Fig. \ref{fig:github}, the patch committer mentioned that the commit fixed the \emph{null pointer deference} vulnerability type.
Code segments can offer implicit information.
For example, the \href{https://github.com/apache/struts/commit/6e87474f9ad0549f07dd2c37d50a9ccd0977c6e}{\textcolor{blue}{Commit:0977c6e}} has an added code ``Pattern allowedNamespaceNames = Pattern.compile("[a-zA-Z0-9.$\backslash$\_/$\backslash\backslash$-]*");'', which is related to input validation.
In addition, there are also commits whose messages or codes are short and obscure, i.e., \href{https://github.com/krasCGQ/linux-vk/commit/6994eefb0053799d2e07cd140df6c2ea106c41ee}{\textcolor{blue}{Commit:06c41ee}} has only commit message while \href{https://github.com/apache/struts/commit/6e87474f9ad0549f07dd2c37d50a9ccd0977c6e}{\textcolor{blue}{Commit:0977c6e}} has only code changes.
Hence, using both types of information can provide more hints to the classifier and improve the chance for accurate prediction. 

%\zc{Compared with using only added and deleted code segments, using all code segments get the best AUC performance. ??0.67 vs 0.62? why say all code segments is better? you mean commit message and all code segments?}
Compared with using commit message with only added and deleted code segments, using commit message with all code segments get the best AUC performance.
%\zc{We observed extra unchanged code segments within the commit can provide important information including reference relationship among APIs and informative comments showing the functionality of APIs or variables. ??then why only all code 0.62 worse than only added+deleted?}
We observed extra unchanged code segments within the commit can provide important information including reference relationships among APIs and informative comments showing the functionality of APIs or variables.
For instance, in \href{https://github.com/apache/tomcat/commit/3aa8f28db7efb311cdd1b6fe15a9cd3b167a2222#}{\textcolor{blue}{Commit:67a2222}}, unchanged code comment indicates the changed code is about validation of the directory path name used for uploaded file storage, which is highly related to the input validation.

\vspace{1mm}
\noindent\fbox{\begin{minipage}{8.6cm} Using both commit messages and all code segments can achieve the best performance in the CodeBERT model. \end{minipage}}\\

\subsection{RQ2: Accuracy of Explainable Silent Dependency Alert Generation}
Similar to RQ1, we investigate both the effectiveness of different models and combinations of inputs on explainable silent dependency alert generation.
From Table \ref{tab:sta_key_aspect}, among the 8,946 vulnerable patching commits, 5,545 commits are for vulnerability type generation, 5,178 commits are for root cause generation, 5,752 commits are for attack vector generation, and 8,585 commits are for impact generation.
{We perform the similar 5-fold cross-validation experiments.}
As our task is an abstractive summarization task, we use Rouge-1, Rouge-2, and Rouge-L \cite{lin2004rouge} which are widely used in the text summarization domain as the evaluation metrics for alert generation.
%Rouge score is widely accepted in the text summarization domain as an evaluation standard that is convenient and effective. 
%\zc{... explain briefly how to compute rouge ...}
%Rouge-1 and Rouge-2 are calculated by the rates of 1-gram and 2-gram tokens that appear both in the prediction and the ground truth to all of the tokens in the ground truth.
%Rouge-L is calculated by the rates of longest common sub-sequence to all of the tokens in the prediction and the ground truth. 
%Due to the use of 2-grams and the largest common sub-sequence, Rouge-2 and Rouge-L can take the order of generated words into account.

%\zc{The best input for classification and generation are different. So do we use different input for classification and generation?}

\subsubsection{Effectiveness of Different Models}
\label{sec:gen_model}

\begin{table*}
	\centering
	%\vspace{1mm}
	\caption{Results of Different Models of Aspect Generator}
	%\vspace{-1mm}
	\label{tab:result_gen_model}
	\setlength{\tabcolsep}{2.5pt}{
		\begin{tabular}{|c|c|c|c|c|c|c|c|c|c|c|c|c|}
			\hline
			&\multicolumn{3}{c|}{Vulnerability Type}&\multicolumn{3}{c|}{Root Cause}&\multicolumn{3}{c|}{Attack Vector}&\multicolumn{3}{c|}{Impact}\\
			%&\multicolumn{3}{c|}{}&\multicolumn{3}{c|}{}&\multicolumn{3}{c|}{}&\multicolumn{3}{c|}{}\\
			\hline
			
			&Rouge-1&Rouge-2&Rouge-L&Rouge-1&Rouge-2&Rouge-L&Rouge-1&Rouge-2&Rouge-L&Rouge-1&Rouge-2&Rouge-L\\
			\hline
			
			BART&\textbf{46.279}&\textbf{41.940}&\textbf{46.148}&31.943&23.639&31.403&34.585&23.563&34.430&34.647&27.292&34.180\\
			\hline
			
			CodeBERT&45.279&41.717&45.253&\textbf{32.770}&\textbf{23.881}&\textbf{32.017}&\textbf{36.134}&\textbf{25.606}&\textbf{36.016}&\textbf{35.217}&\textbf{27.700}&\textbf{34.669}\\
			\hline
			
			 BART-CodeBERT&\multirow{1}{*}{38.896}&\multirow{1}{*}{30.382}&\multirow{1}{*}{38.848}&\multirow{1}{*}{30.244}&\multirow{1}{*}{21.933}&\multirow{1}{*}{29.707}&\multirow{1}{*}{35.468}&\multirow{1}{*}{24.850}&\multirow{1}{*}{35.477}&\multirow{1}{*}{31.925}&\multirow{1}{*}{23.902}&\multirow{1}{*}{31.443}\\
			\hline
			
			Transformer  &\multirow{1}{*}{34.502}&\multirow{1}{*}{30.281}&\multirow{1}{*}{34.576}&\multirow{1}{*}{29.262}&\multirow{1}{*}{20.555}&\multirow{1}{*}{28.615}&\multirow{1}{*}{34.019}&\multirow{1}{*}{22.688}&\multirow{1}{*}{33.828}&\multirow{1}{*}{30.612}&\multirow{1}{*}{21.836}&\multirow{1}{*}{29.983}\\
			%Transformer&&&&&&&&&&&&\\
			\hline
			
			LSTM&11.634&9.333&11.706&21.105&13.478&20.777&35.435&24.945&35.292&20.373&14.405&20.078\\
			\hline
	\end{tabular}}
	\vspace{-3mm}
\end{table*}

%Through the experiment of Section \ref{sec:gen_input}, the input using commit message and added and deleted code segments works the best overall, so we use this input combination to validate effectiveness of different models.

Similar to Section \ref{sec:class_model}, we use the most comprehensive input (i.e., commit message and all code segments) to investigate the model's effectiveness.
Table \ref{tab:result_gen_model} summarizes the Rouge scores of 4 aspects generated by different models.
From Table \ref{tab:result_gen_model}, pre-trained models (BART, CodeBERT, and BART-CodeBERT) are much better than non-pre-trained models (Transformer and LSTM) in all four key aspects.
%There can be a difference of around 34 Rouge scores between them.
This demonstrates the advantages of model pre-training.
%models have great advantages in understanding commit contents than untrained model. 
Compared with the other three aspects, attach vector has the smallest performance gap among the five models.
%betweeThere is no big gap in Rouge scores of attack vector.
This is because the attack vectors of many CVEs are extracted from the IBM X-Force data, which show only the general attack vector information such as ``sending specially-crafted files''.
This makes the generation of attack vector easier than the other three aspects which have more unique information for different vulnerabilities.
%so all five models perform similarly.
%With the decreasing difficulty of attack vector generation, all five models perform similarly.
%Taken together, the pre-trained models are better than the untrained models in vulnerability key aspect generation.

Through the manual review, LSTM and non-pre-trained Transformer have poor results as they cannot handle the details of code contents well, and are more susceptible to interference from the noises.
For example, for vulnerability type generation, in \href{https://github.com/phpmyadmin/phpmyadmin/commit/89fbcd7c39e6b3979cdb2f64aa4cd5f4db27eaad}{\textcolor{blue}{Commit:b27eaad}}, the changed code is in PHP, and there are many \emph{cross-site scripting} vulnerabilities in the training set that come from PHP, so non-pre-trained models tend to generate \emph{cross-site scripting} as the vulnerability type.
However, \href{https://github.com/phpmyadmin/phpmyadmin/commit/89fbcd7c39e6b3979cdb2f64aa4cd5f4db27eaad}{\textcolor{blue}{Commit:b27eaad}} fixed the SQL parts in the PHP code, and the correct vulnerability type is \emph{SQL injection}, which is ignored by the non-pre-trained models.
For the generation of root cause, in \href{https://github.com/symphonycms/symphonycms/commit/1ace6b31867cc83267b3550686271c9c65ac3ec0}{\textcolor{blue}{Commit:5ac3ec0}}, the changed code is about \emph{input validation}, but because it involves some modifications to the array variables, the non-pre-trained models incorrectly judge it as \emph{improper bounds checking}. %\zc{\emph{improper bounds checking} ??isn't it about input validation?}. 
For the attack vector, the non-pre-trained models judge the \href{https://github.com/ljalves/linux_media/commit/946e51f2bf37f1656916eb75bd0742ba33983c28}{\textcolor{blue}{Commit:3983c28}}'s attack vector as \emph{via a name parameter}, because the code changes involve file name variables.
Nevertheless, the name variables are not related to the vulnerability, and the true reason of the vulnerability is file deadlock related to the wrong logic.
%The correct attack vector should be \emph{via a crafted application}.
For impact,  the non-pre-trained models judge the \href{https://github.com/ravidholakia/symfony/commit/4fb975281634b8d49ebf013af9e502e67c28816b}{\textcolor{blue}{Commit:c28816b}} 's impact as \emph{steal cookie files}, while the true impact is \emph{file deletion} which is similar to the file stolen impact.

Among three pre-trained encoder-decoder models, CodeBERT gets the best outcomes in {the three out of four tasks}, followed by BART.
This is because the pre-trained code contents play a decisive role, providing code knowledge that cannot be offered by the pre-trained data.
The results of the BART-CodeBERT model are worse than the BART and CodeBERT.
%The pre-trained BART encoders and decoders are not compatible with pre-trained CodeBERT encoders, leading the score decreasing.

\vspace{1mm}
\noindent\fbox{\begin{minipage}{8.6cm} Pre-trained models can offer huge advantages for alert generation. CodeBERT generator achieves the best overall results due to its code-specific pre-training. \end{minipage}}\\

\subsubsection{Effectiveness of Different Input Combinations}
\label{sec:gen_input}

\begin{table*}
	\centering
	\vspace{1mm}
	\caption{Results of CodeBERT-based Aspect Generator for Different Types of Inputs}
	%\vspace{-1mm}
	\label{tab:result_gen_input}
	\setlength{\tabcolsep}{2.5pt}{
		\begin{tabular}{|c|c|c|c|c|c|c|c|c|c|c|c|c|}
			\hline
			&\multicolumn{3}{c|}{Vulnerability Type}&\multicolumn{3}{c|}{Root Cause}&\multicolumn{3}{c|}{Attack Vector}&\multicolumn{3}{c|}{\multirow{1}{*}{Impact}}\\
			%&\multicolumn{3}{c|}{}&\multicolumn{3}{c|}{Cause}&\multicolumn{3}{c|}{Vector}&\multicolumn{3}{c|}{}\\
			\hline
			
			&Rouge-1&Rouge-2&Rouge-L&Rouge-1&Rouge-2&Rouge-L&Rouge-1&Rouge-2&Rouge-L&Rouge-1&Rouge-2&Rouge-L\\
			\hline
			
			Commit Message&43.875&40.049&43.948&28.236&20.267&27.777&32.928&22.186&32.895&34.497&26.162&33.678\\
			\hline
			
			Added \& Deleted&\multirow{2}{*}{37.679}&\multirow{2}{*}{34.141}&\multirow{2}{*}{37.688}&\multirow{2}{*}{30.420}&\multirow{2}{*}{22.291}&\multirow{2}{*}{29.675}&\multirow{2}{*}{35.458}&\multirow{2}{*}{25.121}&\multirow{2}{*}{35.318}&\multirow{2}{*}{33.032}&\multirow{2}{*}{25.511}&\multirow{2}{*}{32.456}\\
			Code Segments&&&&&&&&&&&&\\
			\hline
			
			All Code&\multirow{2}{*}{38.560}&\multirow{2}{*}{34.866}&\multirow{2}{*}{38.471}&\multirow{2}{*}{29.855}&\multirow{2}{*}{21.282}&\multirow{2}{*}{29.208}&\multirow{2}{*}{36.155}&\multirow{2}{*}{25.722}&\multirow{2}{*}{36.074}&\multirow{2}{*}{31.033}&\multirow{2}{*}{24.170}&\multirow{2}{*}{30.709}\\
			Segments&&&&&&&&&&&&\\
			\hline
			
			Commit Message \& &\multirow{3}{*}{\textbf{45.387}}&\multirow{3}{*}{\textbf{41.812}}&\multirow{3}{*}{\textbf{45.403}}&\multirow{3}{*}{\textbf{32.035}}&\multirow{3}{*}{23.376}&\multirow{3}{*}{\textbf{31.436}}&\multirow{3}{*}{\textbf{36.325}}&\multirow{3}{*}{25.500}&\multirow{3}{*}{\textbf{36.145}}&\multirow{3}{*}{\textbf{37.670}}&\multirow{3}{*}{\textbf{30.112}}&\multirow{3}{*}{\textbf{36.904}}\\
			Added \& Deleted&&&&&&&&&&&&\\
			Code Segments&&&&&&&&&&&&\\
			\hline
			
			Commit Message \& &\multirow{2}{*}{45.280}&\multirow{2}{*}{41.717}&\multirow{2}{*}{45.253}&\multirow{2}{*}{31.943}&\multirow{2}{*}{\textbf{23.640}}&\multirow{2}{*}{31.404}&\multirow{2}{*}{36.135}&\multirow{2}{*}{\textbf{25.607}}&\multirow{2}{*}{36.016}&\multirow{2}{*}{34.647}&\multirow{2}{*}{27.292}&\multirow{2}{*}{34.181}\\
			All Code Segments&&&&&&&&&&&&\\
			\hline
	\end{tabular}}
\end{table*}	

Considering Section \ref{sec:gen_model}, we choose the CodeBERT encoder-decoder model as the test target.
Table \ref{tab:result_gen_input} summarizes the results of CodeBERT for different inputs.
It shows that inputs with both commit messages and code contents have much better results, indicating both commit messages and code contents are important to the generation.
In some patches, the aspect appears directly in the commit messages.
For example, \href{https://github.com/ADVAN-ELAA-8QM-PRC1/platform-external-ImageMagick/commit/7b1cf5784b5bcd85aa9293ecf56769f68c037231}{\textcolor{blue}{Commit:c037231}} contains a message ``Fixed out of bounds error in SpliceImage'', which directly shows the vulnerability type \emph{out of bounds error}.
The code segments can also be clues for generation.
For example, \href{https://github.com/mcdonough-john82/OpenCV/commit/f0fb665407a1162153930de1ab5a048a4f3d60f9}{\textcolor{blue}{Commit:f3d60f9}} has a large number of image size calculations and exception handling, making the model predict the vulnerability type as the \emph{buffer overflow}.
In general, commit messages are more useful for generating vulnerability type and impact, while code contents are more useful for generating root cause and attack vector.
%leading to better results of using pure commit message in generation of vulnerability type and impact compared with using pure code contents.

Using commit message and added and deleted code segments have slightly better results than using commit message and all code segments in most Rouge metrics.
The difference between the two inputs is whether having unchanged code segments.
Our results suggest that unchanged code segments may interfere with the model's attention to important code changes related to key aspects, although unchanged code segments are beneficial for vulnerability patch classification.
%the training, making model fails to pay attention to the most important information.
%Overall, from the table, the help of unchanged code segments in generation is not obvious in most of the generations.

\vspace{1mm}
\noindent\fbox{\begin{minipage}{8.6cm} Different input combinations have different helpfulness to the generation. Overall, the input of commit message and added and deleted codes produces the better outcomes. \end{minipage}}\\

\subsection{RQ3: Usefulness of Explainable Silent Dependency Alert Prediction}
\label{sec:usefulness}

We conduct a small-scale user study to investigate the usefulness of our explainable silent dependency alert prediction, compared with just binary silent dependency alert prediction.

\subsubsection{User Study Design}
Table \ref{tab:cha3-rq3questions} lists the six commits used as tasks in our study, the related products, the correctness of binary silent dependency alert prediction, and the generated explainable vulnerability key aspects. 
We randomly select vulnerable and non-vulnerable commits as the study task. 
For each task, we provide the whole commit information including links, messages, and code segments to the participants.
All the participants can click the commit links to see all related files and file changes.
In addition, we provide the vulnerability patch classification results and the generated explainable key aspects as hints.
%Based on AI classification, explanation and own experience, 
We ask the participants to answer the question ``is this a patching commit for vulnerability or not'' to see whether the extra explainable information may affect the participants' decision and performance.

In the real application of a prediction approach like ours, precision is more important than recall.
%The traditional models suffer from high ratios of false positive results, wasting security experts' time to validate the irrelevant risks.
The false positive vulnerability patches are inevitable, no matter how accurate the prediction model is.
Such false positive predictions will waste security analysts' time validating the irrelevant risks.
For false positive patches, the generated key aspects are irrelevant.
Studies on XAI~\cite{nourani2021anchoring, nourani2020don} show irrelevant explanations to AI decisions are even worse than no explanations.
We are interested in investigating how irrelevant explanations affect vulnerability patch analysis.
Inspired by such context, we focus on true positive and false positive results in this study to evaluate the practical value of our approach. 
The selected tasks include 4 vulnerable commits (T1-T4) (true positives) and 2 non-vulnerable commits (T5-T6) (false positives) that are all predicted as vulnerable commits by the classifier.
%Based on analysis, we separate selected tasks into 3 difficulty classes, with 2 easy tasks, 2 medium-level tasks and 2 hard tasks.
Based on our experiments in RQ1 and RQ2, we use commit message and all code segments as the inputs with trained CodeBERT classifier to generate binary silent dependency alert prediction, and we use commit message, added and deleted code segments as the inputs with trained CodeBERT generator to generate vulnerability key aspects for the six commits.

\begin{table*}
	\centering
	\caption{Six GitHub Commits in Our User Study}
	\vspace{-1mm}
	\label{tab:cha3-rq3questions}
	\setlength{\tabcolsep}{2.5pt}{
	\begin{tabular}{|c|c|c|>{\everypar\expandafter{\the\everypar{\dohang}\raggedright\arraybackslash}\arraybackslash}L{.76\textwidth}|}
		\hline
		&\textbf{Product}&\textbf{Correctness}&\multicolumn{1}{c|}{\textbf{Generated Vulnerability Key Aspect Explanations}} \\
		\hline
		
		\multirow{1}{*}{\href{https://github.com/mtavella-netspot/moodle/commit/ce5a785b0962c3c94c7a7b0d36176482d21db95d}{\textcolor{blue}{T1}}}&\multirow{1}{*}{Moodle}&True&This is patching for \textcolor{mygreen}{cross-site scripting vulnerability}, the root cause is \textcolor{mybrown}{improper validation of user-supplied input}, attacker can exploit by \textcolor{mylightblue}{sending a specially-crafted request} to \textcolor{myblue}{steal the victim's cookie-based authentication credentials}.\\

		\multirow{1}{*}{\href{https://github.com/YashTechnoBoy/phpmyadmin/commit/015c404038c44279d95b6430ee5a0dddc97691ec}{\textcolor{blue}{T2}}}&\multirow{1}{*}{Phpmyadmin}&True&This is patching for \textcolor{mygreen}{cross-site scripting (xss) vulnerability}, the root cause is \textcolor{mybrown}{improper input validation}, attacker can exploit \textcolor{mylightblue}{via the search parameter} to \textcolor{myblue}{obtain sensitive information}.\\
		
		\multirow{1}{*}{\href{https://github.com/Krylon360/evernote-graphite-web/commit/4a9f98647be279a39a982bd94922fdec710b0b3f}{\textcolor{blue}{T3}}}&\multirow{1}{*}{Graphite-web}&True&This is patching for \textcolor{mygreen}{code execution vulnerability}, the root cause is \textcolor{mybrown}{improper validation of user-supplied input}, attacker can exploit by \textcolor{mylightblue}{sending a specially-crafted request} to \textcolor{myblue}{execute arbitrary code on the system}.\\
		
		\multirow{1}{*}{\href{https://github.com/YashTechnoBoy/phpmyadmin/commit/19df63b0365621427697edc185ff7c9c5707c523}{\textcolor{blue}{T4}}}&\multirow{1}{*}{Phpmyadmin}&True&This is patching for \textcolor{mygreen}{sql injection vulnerability}, the root cause is \textcolor{mybrown}{is improper validation of user-supplied input}, attacker can exploit \textcolor{mylightblue}{via the search parameter} to \textcolor{myblue}{modify or delete information in the back-end database}.\\

		\multirow{1}{*}{\href{https://github.com/joansmith1/okhttp/commit/d9e3e7dd3361cf097e5d5fc4db76f3c89058f82b}{\textcolor{blue}{T5}}}&\multirow{1}{*}{Okhttp}&False&This is patching for \textcolor{mygreen}{remote code execution vulnerability}, the root cause is \textcolor{mybrown}{improper validation of user-supplied input}, attacker can exploit by \textcolor{mylightblue}{sending a specially-crafted request} to \textcolor{myblue}{the application to crash}.\\
		
		\multirow{1}{*}{\href{https://github.com/openssl/openssl/commit/63b996e752ac698186c38177232280e6515d571b}{\textcolor{blue}{T6}}}&\multirow{1}{*}{OpenSSL}&False&This is patching for \textcolor{mygreen}{information leak vulnerability}, the root cause is \textcolor{mybrown}{the failure to initialize certain structures}, attacker can exploit by \textcolor{mylightblue}{sending a specially-crafted request} to \textcolor{myblue}{obtain sensitive information}.\\
		
		\hline
	\end{tabular}}
	\vspace{-3mm}
\end{table*}

%We recruit 10 participants from our partner's company %{here I have concern as we add dehai as the author, it will be obvious our partner is Huawei...} who are responsible for tracking and analyzing the vulnerabilities of OSS libraries. 
%These participants have the expertise for the tasks and their feedback will help us understand the usefulness of our approach in practice.
%They are exact security testers in the commercial companies that can reflect the real situation of commercial needs. 
We recruit 10 students {of whom six are male and four are female} from software security domain as participants from our schools.
{All the participants are concentrated in 18-24 years old.}
They all have {one to three years experience working in the organization as the developer and the tester with} basic knowledge of tracking and analyzing the vulnerabilities of OSS libraries, which are similar to the testers in many commercial companies, so we can imitate real situations in the testing process.
We confirm none of the participants know the tasks' answers. 
Based on the pre-experiment survey of the participants' programming and security analysis experience, we randomly assign them into two comparable groups. 
The experimental group uses our explainable silent dependency alert prediction offering the classification results and a short vulnerability description of vulnerability key aspects. 
The control group can only see the classification results.
%Both of groups can view the code files and changes included in the commits, and the other files included in the corresponding projects.
Before the study, we provide a 10-minutes explanation for our classification and explanation results to both the experimental group and the control group. 
To ensure fairness and avoid bias, we only explain the meaning of classification and explanation. 
%but do not provide goals or technical details of our work to make sure the participants do not have preference for the generation results.
However, we exclude goals or technical details of our work to make sure the participants have an insufficient preference for the generation results.
We confirm all the participants understand the question and predictions through an easy pilot task.
The participants are given 10 minutes for each task.
If the participants complete the task, they can submit their answers before the timeout. 
We record the task completion time, the participants’ ratings of task difficulties, and the usefulness of explainable key aspects for analysis using a 5-point Likert scale for each task. 
If the participants cannot complete the task before timeout, we consider the current task is failed and require the participants to start the next task. 
We record participants' screens for the post-experiment analysis of their task behaviours.

\subsubsection{User Study Results}

\begin{table}%[htbp]
	\centering
	\caption{Performance Comparison for Each Task between Two Groups}
	\vspace{-1mm}
	\label{tab:cha7-performance}
	\begin{tabular}{|c|c|c|c|c|}
		\hline
		
		\multirow{2}{*}{Index}&\multicolumn{2}{c|}{Experimental Group}&\multicolumn{2}{c|}{Control Group}\\
		\cline{2-5}
		&AveTime (s)&\# Correct&AveTime (s)&\# Correct\\
		%	&(sec.)&&(sec.)&\\
		\hline
		T1&121.2&5&63.2&5\\
		\hline
		
		T2&110.6&3&86.8&1\\
		\hline
		T3&134.6&5&189.2&4\\
		\hline
		
		T4&59.0&4&168.6&4\\
		\hline
		
		T5&125.6&2&121.0&3\\
		\hline
		T6&170.0&4&157.2&2\\
		\hline
		Ave$\pm$Stdev&\multirow{1}{*}{120.2$\pm$36.2}&&\multirow{1}{*}{131.0$\pm$49.3}&\\
		%stddev&&&&\\
		\hline
	\end{tabular}
	\vspace{-5mm}
\end{table}

Table \ref{tab:cha7-performance} shows the average task completion time (AveTime) and the correctness (\# Correct) for each task between the experimental group and control group.
Compared with the control group knowing only the binary classification outcomes, the experimental group with explainable information has shorter task completion time (120.2$\pm$36.2 seconds versus 131.0$\pm$49.3 seconds) and higher correct judgements (23 correct answers versus 19 correct answers).
Experimental group spent more time on tasks T1, T2, T5, and T6.
From recorded videos, experimental group members have to read and consider the correctness of explainable information, especially for the false positive predictions (T5-T6), which adds extra time to the tasks.
Another reason we found is both the T1 and T2 are fixes of cross-site scripting vulnerability that try to add validation to the POST parameters of URI, which is easy for the control group to make decision.
However, after reviewing the explainable information, the experimental group members make more correct decisions to each task, which proves the extra spent time is worth it.
Under the best situation (T4), the experimental group uses 65\% less time than the control group.
The above results indicate with help of the explainable information, participants can recognize silent dependency alerts quicker with a higher correctness rate.

Considering false positive tasks (T5 \& T6), the experimental group has to spend more time making their decision, while the overall correctness does not increase.
%is similar between the two groups.
%, while keeping almost similar correctness to the control group.
By reviewing the videos, we find that as the aspect explanation is irrelevant to the vulnerability, this seems to confuse the experimental participants.
They need more time to deny the prediction.
This outcome suggests that the current form of vulnerability explainable information is not helpful to relieve the issue of false positive prediction, but can better support the true positive prediction and help the analysts to make correct decisions easily.
Such results are consistent with the observation of the XAI study in \cite{nourani2020don}, which shows that irrelevant explanations are worse than no explanations to help users understand AI decisions.

\begin{comment}
	\begin{figure}
	\centering
	\includegraphics[scale=0.6]{images/difficult.png}
	\caption{Average Task Difficulty Labelled by Two Groups}
	\label{fig:difficult}
	\end{figure}

	\begin{figure}
	\centering
	\includegraphics[scale=0.6]{images/usefulness.png}
	\caption{Average  Explainable Generation Usefulness Labelled by Experimental Group}
	\label{fig:usefulness}
	\end{figure}
	\vspace{-3mm}
\end{comment}

\begin{figure}%[htbp] 
	\centering
	\subfigure[Average Task Difficulty Labelled by Two Groups]{\includegraphics[width=.23\textwidth]{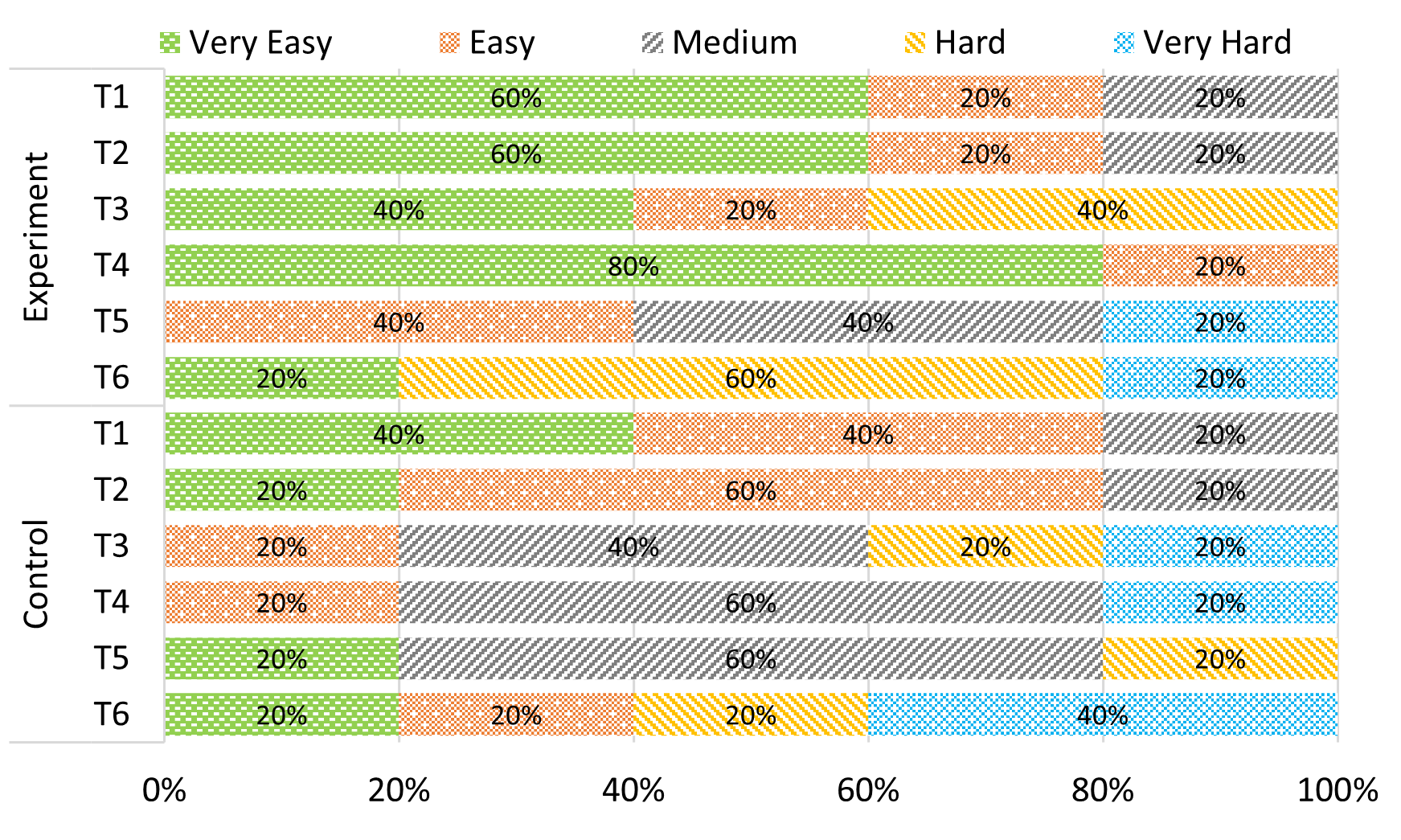}
		\label{fig:difficult}	
	}
	\hfil
	\subfigure[Average  Explainable Generation Usefulness Labelled by Experimental Group]{\includegraphics[width=.22\textwidth]{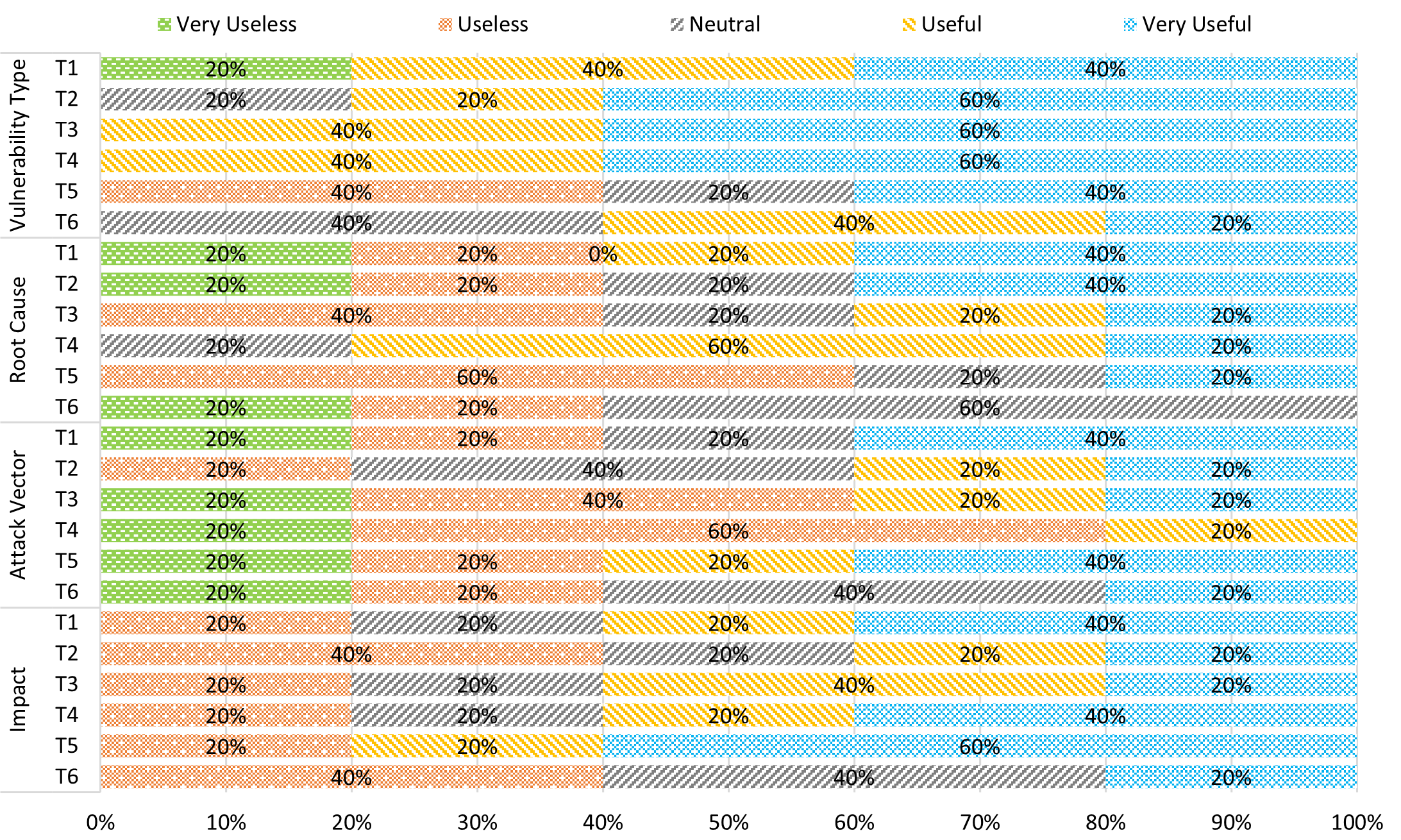}
		\label{fig:usefulness}	
	} 
	\vspace{-2mm}
	\caption{5 Likert Scale of Difficulty and Usefulness Marks}
	%\vspace{-7mm}
	%  \textcolor{blue}{Siri: Tick label is hard to see. Try to enlarge it. You can use narrow bar if necessary.}
	\label{fig:useful_diff}
	\vspace{-5mm}
\end{figure}

%\zc{http://daydreamingnumbers.com/blog/4-ways-to-visualize-likert-scales/ Use this type of visulization of likert results. Ensure your discussion with the messages in the new visulaization.}

Fig. \ref{fig:difficult} shows the task difficulty ratings by two groups.
For all tasks with true positive predictions (T1-T4), the experimental group gives more very easy/easy comments.
%The difference of difficulty level can be as high as 2 points between the two groups.
This indicates that the explainable information can help to relieve the difficulties of deciding the true silent dependency alerts. 
%and this advantage becomes obvious to the medium-level difficulty task including T3 and T4.
%Being consistent with Table \ref{tab:cha7-performance}, experimental participants spend less time but have higher correctness in the task T3 and T4.
In contrast, tasks with false positive predictions (T5-T6) obtain more difficulty ratings including hard/very hard comments and extreme options from the experimental group, which indicates the experimental group feels obstacles in making decisions.
The generated explanation is irrelevant to the prediction, which can affect the analysis and decision by the participants.
 
Fig. \ref{fig:usefulness} shows the evaluation of the usefulness of each generated key aspect in each task.
Overall, the experimental group believes the generated key aspects are useful, especially for the true positive predictions (T1-T4).
Vulnerability type and impact are rated to be more useful than the other two aspects, as 60\%-80\% of comments reflect the two aspects are useful/very useful in the tasks.
As shown in Table \ref{tab:cha3-rq3questions}, generated vulnerability type and impact are more diverse than the two other aspects due to the semantic feature of NVD and IBM X-Force.
As such, vulnerability type and impact can provide more contribution to recognizing silent dependency alerts.
Compared with the tasks with true positive predictions (T1-T4), explainable key aspects received lower usefulness and more extreme options in the tasks with false positive predictions (T5-T6), indicating the helpfulness of generation in false positive predictions is not obvious.

\vspace{1mm}
\noindent\fbox{\begin{minipage}{8.6cm} The participants, with help of explainable aspects, recognize the true positive silent dependency alert quicker and more accurate. In the contrast, the false positive explainable silent dependency alert prediction could mislead the participants to spend more time analyzing incorrect model outcomes. \end{minipage}}\\

	\section{Threats to the Validity}
%\textbf{Internal validity:} 
%\zc{... some discussion on the data collection process and data validity ...}
%We rely on external resources including Serena et al.'s project \cite{Ponta@2019} and Liu et al.'s project \cite{Liu@icse2020} as ground truth for our silent dependency alert prediction without validating their correctness.
%As their data is collected manually, it is inevitable to contain human errors.
%Nevertheless, even the official NVD database also contains mistakes \cite{nguyen2013reliability}, and we believe the most of the data is correct as the data is collected by the security experts, so the overall impact on the model should be tiny.

\textbf{Internal validity:} We evaluate the effect of our prediction on true and false positive predictions.
Although our approach has very high accuracy, the negative predictions and the effect of aspect explanations need further investigation. 
However, there is no one-for-all method to solve a problem, and by far, our method achieves the state-of-art results.
Our framework is the first work applying XAI in silent dependency alert prediction, and we hope our work can open the door to enhance the explanation of AI techniques in this domain.

\textbf{External validity:} As we mix the programming language of data, it is difficult to determine the influence of a specific language on the experiments. 
However, due to the complexity of patch data collection, it is difficult to collect enough data for each language.
%Therefore, we mixed the patches of different languages. 
Overall, our dataset is large and diverse, coming from 2,017 repositories, and our approach achieves good prediction results (0.89 AUC scores and 32.8-45.2 Rouge1 scores).
Therefore, our results bear a certain level of guiding significance. 
In the future, we will investigate both language-agnostic and language-specific vulnerability patch analysis. 
%try to manually label different types of data to make the results more comprehensive.

{Our user study only contains 6 questions with 10 participants, which can limit the statistical evidence of our discoveries.
We found no significant difference between the experimental and control groups, as the correctness and completion time between the two groups are close.
Nevertheless, the study requires special expertise, which increase the difficulties for finding suitable participants.
Compared to quantity, we pay more attention to quality and only invite the experienced participants to simulate the real-world situation as much as possible.
We believe that the user study reveals the effectiveness and pain points of explanation in vulnerability analysis. 
Specifically, we show that our explainable vulnerable key aspects help to recognize vulnerable code faster and more accurately (120.2±36.2 vs. 131.0±49.3 seconds and 23 vs. 19 correct answers, respectively), and the false generation will impair user performance, which are coincident to the previous works \cite{nourani2020don}. 
Most importantly, our study is the first work to conduct a user study for vulnerability detection, which can be instructive for the future research in the relevant area.
}

{We only check the effectiveness of true and false positive predictions in the user study, and the impacts of negatives remain unknown. 
False negatives can cause damages as the vulnerability patches are missing.
However, in the real usage scenario, when one must deal with thousands of software products and their dependencies, the developers will be easily overwhelmed by too many false positives.
Hence, the maintaining people prefer to allocate the effort to accurate predictions over filtering false negatives, so they pay more attention to the precision rather than recall.
In this study, we select the most important pain point we believe for better understanding.
We leave the full research of XAI impact to the future works.
}

	\section{Related Work}

%\zc{Missing some highly related works ...}

\textbf{Silent vulnerable patch detection.} 
Much work has been focused on using static, dynamic, or feature-based methods to detect vulnerability dependency alerts \cite{Papagiannis@2011, Viega@2000, Younan04codeinjection} that focus on specific program languages.
%These works can identify vulnerabilities by analyzing code abstract syntax tree, or compare code similarity.
%However, these methods can only be applied to specific program languages and limited vulnerability types.
Recently, machine learning including LSTM and CodeBERT has been adopted for its analytic capability \cite{Zhou@2017, zhou2021spi, zhou2021finding, Zhen@2020,Chen@2020,Ramsauer@2020,Sabetta@2018}.
%Zhou and Sharma \cite{Zhou@2017} use pre-defined rule-based method to filter the candidates of vulnerability patching commits, and use feature engineering with the combination of multiple machine leaning models to judge the results.
%Zhou et al.\cite{zhou2021spi} uses add and deleted code segments as the inputs with traditional LSTM classifier to determine the vulnerability patching commits.
%The are many similar works \cite{Zhen@2020,Chen@2020,Ramsauer@2020,Sabetta@2018}.
%These works all use pre-Transformer machine learning models, and the prediction accuracy is not high.
Zhou et al.' s work \cite{zhou2021finding} is the most similar work to ours, using CodeBERT model to detect silent dependency alerts, but they exclude provide explanations and validate the helpfulness. 
%However, none of these tools provide prediction explanations.
%\zc{... how about Jiayuan's ASE work ...}
%Due to low accuracy, their results are not very convincing. 
%Developers have to manually check the vulnerable code lines and determine corresponding vulnerability key aspects to understand the vulnerability in the code. 
%Our experiments show explainable key vulnerability aspects can help the developer determine the validity of the vulnerability patch prediction.

\begin{comment}
\textbf{Vulnerability key aspect extraction. } 
%There are also some works similar to ours, which also implement the extraction of the vulnerability key aspect. 
Sun et al. \cite{sun2021generating} apply BERT-based NER and QA techniques to extract short and long key aspects from vulnerability textual descriptions for CVE generation from ExpoitDB posts.
Dong et al. \cite{Dong@usenix2019} also use NER techniques to extract vulnerability key aspects from different security forms to understand the vulnerability key aspect inconsistency across different security platforms.
%However, all these works use extractive summarization of vulnerability key aspect extraction from the textual vulnerability descriptions, and cannot deal with OSS commits that lack necessary vulnerability aspect information.
However, these works are mainly for the exploration of CVE report issues, such as the inconsistency of the vulnerability key aspect of different websites, but our target is to predict potential OSS vulnerability patching commits.
\end{comment}

\textbf{Explainable AI.} 
Explainable AI has the potential to alleviate the significant concerns about black-box AI models.
Researchers find that the true explanation can enhance the performance of AI models, while the wrong explanation can mislead users \cite{weld2019challenge, nourani2020don, nourani2021anchoring}.
%Weld et al. find that the extra explainable information can help to enhance the performance of AI models \cite{weld2019challenge}. 
%Nourani et al. \cite{nourani2020don} focus on the effectiveness of Explainable AI on video activity recognition. 
%They find that the correct explainable information is helpful, but the wrong explainable message can mislead users.
%Nourani et al. \cite{nourani2021anchoring} examine the user reliance on Explainable AI systems, and show that both over-reliance and underestimation can damage the effect of AI models.
Explainable AI can explain AI results in various forms including visual explanations \cite{simonyan2013deep, zeiler2014visualizing}, verbal explanations \cite{berkovsky2017recommend, lakkaraju2016interpretable}, and analytic explanations \cite{hohman2018visual}.
%Because of its effectiveness, Explainable AI has been used in different domains including explainable recommendation systems \cite{wang2018tem, cheng2019mmalfm}, classification systems \cite{ribeiro2016should} and activity recognition systems \cite{roy2019explainable, atzmueller2018explicative}.
%None of work above uses Explainable AI in AI-based software vulnerability analysis.

{
Tantithamthavorn et al.'s work provides basic knowledge of XAI, and proposes explainable AI techniques to address the software defects \cite{tantithamthavorn2021explainable}.
Jiarpakdee et al.'s work evaluates three popular XAI tools to figure out the usefulness of explanation in the software defect detection \cite{jiarpakdee2020empirical}. 
Li et al.'s work uses Program Dependency Graph to interpret results of a vulnerability
detector \cite{li2021vulnerability}.
Similarly, Fu et al.'s work tries to provide explanation of vulnerability detection by locating the questionable functions and code lines \cite{9796256}.
Their explanation focuses on code lines related to the vulnerability, while ours is external vulnerability properties explained in natural language, hence the two types of explanations complement each other.
There are many other works prove the necessity of XAI in the software domain \cite{jiarpakdee2021practitioners, wattanakriengkrai2020predicting, pornprasit2021jitline}.
}
%In this work, we evaluate the effectiveness of verbal explanations in the vulnerable dependency alert classification problem, and find its helpfulness as well as limitations.
%\zc{Do you cover Tien's FSE 2021 paper Vulnerability Detection with Fine Grained Interpretation. I think their explanation is some code lines related to the vulnerability, while ours is external vulnerability properties explained in natural language. The two types of explanations complement each other.}

\textbf{Pre-trained code language models and their applications.}
%\zc{As the core is based on CodeBert, we need to summarize CodeBert-based downstream tasks. See a list here %\url{https://paperswithcode.com/method/codebert#:~:text=CodeBERT}. I think it includes some work on vulnerability analysis.}
Feng et al. design code-related pre-training tasks to propose a code-sensitive BERT named CodeBERT that acquires state-of-art results in many code-related tasks \cite{feng2020codebert} and has been widely used in domains including question answering, binary vulnerability detection, transformation, and data augmentation \cite{huang2021cosqa, zhou2021finding, nguyen2021regvd, shi2022enhancing, goel2022cross}.
Our work extends the usefulness of CodeBERT to explain a silent dependency alert domain, hoping to enhance the security of software supply chains. 
	
	\section{Conclusion}

Driven by the real needs, we proposed a framework for automatically identifying silent dependency alerts with an explanation including vulnerability type, root cause, attack vector, and impact.
We collected vulnerable commits from various data sources and use BERT-based NER and QA models to extract vulnerability key aspects.
By using a CodeBERT-based generator and classifier, we achieved silent dependency alert prediction.
%From Snyk databases, NVD reference lists, Serena et al.'s projects and Liu's et al.'s projects, we collected thousands of OSS vulnerability patching commits with correspond CVE ids.
%By using GitHub API and BERT NER and QA methods, we extracted commit messages, code changes and vulnerability key aspects from different GitHub projects, NVD database and IBM X-Force database.
%With a cross-model self-attention layer, our model can combine different pre-trained Transformer-based model without changing the input dimensions.
Our experiments proved both commit message and code segments are useful for the generation and classification, and CodeBERT achieves the best results. 
%Currently, our model suffers from the mixture of code languages due to lack of the training data.
In the future, we will label commit patches for specific code languages to further validate model usefulness.
%We also proved that pre-train Transformer models can be better than the untrained models.
%From the evaluation, our generator can achieve around 25 to 45 scores for Rouge-1, Rouge-2 and Rouge-L metrics, and achieve 75\%-85\% accuracies.
%Our classifier achieves 0.78 precision, 0.77 recall and 0.78 F1 scores, which is reasonable for application.
%Our user study proves the true positive results of our model can be helpful for recognizing silent dependency alerts.

%Currently, our model suffers from the mixture of different types of code languages due to lack of the training data.
%This prevents us from detecting efficiency of our model for each language types.
%Furthermore, our framework can mislead users with false outcomes.
%In the future, we will manually label commit patching data for specific code languages, so the advantages and disadvantages of the model can be more obvious.
%With more training data, the accuracy of the model can also enhance to avoid more false predictions.
%Besides, we found there is inconsistency issue for vulnerability key aspects among different databases.
%Such issue is also mentioned by Dong et al.'s works \cite{Dong@usenix2019}.
%This issue can affect the generation of vulnerability key aspects as the ground truth is uncertain.
%In the future, we plan to further research the related inconsistency issues, and try to normalize the vulnerability key aspects, so that the generation can have higher quality.

	%\bibliographystyle{ACM-Reference-Format}
	\bibliographystyle{IEEEtran}
	\bibliography{fse2021citelist}

% Generated by IEEEtran.bst, version: 1.14 (2015/08/26)
\begin{thebibliography}{10}
\providecommand{\url}[1]{#1}
\csname url@samestyle\endcsname
\providecommand{\newblock}{\relax}
\providecommand{\bibinfo}[2]{#2}
\providecommand{\BIBentrySTDinterwordspacing}{\spaceskip=0pt\relax}
\providecommand{\BIBentryALTinterwordstretchfactor}{4}
\providecommand{\BIBentryALTinterwordspacing}{\spaceskip=\fontdimen2\font plus
\BIBentryALTinterwordstretchfactor\fontdimen3\font minus
  \fontdimen4\font\relax}
\providecommand{\BIBforeignlanguage}[2]{{%
\expandafter\ifx\csname l@#1\endcsname\relax
\typeout{** WARNING: IEEEtran.bst: No hyphenation pattern has been}%
\typeout{** loaded for the language `#1'. Using the pattern for}%
\typeout{** the default language instead.}%
\else
\language=\csname l@#1\endcsname
\fi
#2}}
\providecommand{\BIBdecl}{\relax}
\BIBdecl

\bibitem{Ponta@2019}
S.~E. Ponta, H.~Plate, A.~Sabetta, M.~Bezzi, and C.~Dangremont, ``A
  manually-curated dataset of fixes to vulnerabilities of open-source
  software,'' in \emph{Proceedings of the 16th International Conference on
  Mining Software Repositories}, ser. MSR '19.\hskip 1em plus 0.5em minus
  0.4em\relax IEEE Press, 2019, p. 383–387.

\bibitem{Ponta@2018}
S.~Ponta, H.~Plate, and A.~Sabetta, ``Beyond metadata: Code-centric and
  usage-based analysis of known vulnerabilities in open-source software,'' 06
  2018.

\bibitem{solarwinds}
I.~Jibilian and K.~Canales, ``{The US is readying sanctions against Russia over
  the SolarWinds cyber attack. Here's a simple explanation of how the massive
  hack happened and why it's such a big deal},''
  \url{https://www.businessinsider.com/solarwinds-hack-explained-government-agencies-cyber-security-2020-12},
  2021, accessed: 2022-04-23.

\bibitem{Log4j}
T.~Hunter and G.~D. Vynck, ``The ‘most serious’ security breach ever is
  unfolding right now. here’s what you need to know.''
  https://www.washingtonpost.com/technology/2021/12/20/log4j-hack-vulnerability-java/,
  Dec. 2021, accessed: 2022-04-23.

\bibitem{Log4j2}
D.~GOODIN, ``The internet’s biggest players are all affected by critical
  log4shell 0-day,''
  https://arstechnica.com/information-technology/2021/12/the-critical-log4shell-zero-day-affects-a-whos-who-of-big-cloud-services/,
  Dec. 2021, accessed: 2022-04-23.

\bibitem{CVE}
{Common Vulnerabilities and Exposures}, \url{https://cve.mitre.org/index.html},
  2022, accessed: 2022-04-23.

\bibitem{NVD}
{National Vulnerability Database}, \url{https://nvd.nist.gov/}, 2022, accessed:
  2022-04-23.

\bibitem{githubadv}
{GitHub Advisory Database}, \url{https://github.com/advisories}, 2022,
  accessed: 2022-04-23.

\bibitem{Dependabot}
Dependabot, \url{https://github.com/dependabot}, 2022, accessed: 2022-04-23.

\bibitem{Ruohonen@2018}
J.~Ruohonen, S.~Rauti, S.~Hyrynsalmi, and V.~Leppänen, ``A case study on
  software vulnerability coordination,'' \emph{Information and Software
  Technology}, vol. 103, pp. 239--257, 2018.

\bibitem{zhou2021finding}
J.~Zhou, M.~Pacheco, Z.~Wan, X.~Xia, D.~Lo, Y.~Wang, and A.~E. Hassan,
  ``Finding a needle in a haystack: Automated mining of silent vulnerability
  fixes,'' in \emph{2021 36th IEEE/ACM International Conference on Automated
  Software Engineering (ASE)}.\hskip 1em plus 0.5em minus 0.4em\relax IEEE,
  2021, pp. 705--716.

\bibitem{Li@2017}
F.~Li and V.~Paxson, \emph{A Large-Scale Empirical Study of Security
  Patches}.\hskip 1em plus 0.5em minus 0.4em\relax New York, NY, USA:
  Association for Computing Machinery, 2017, p. 2201–2215.

\bibitem{Dong@usenix2019}
Y.~Dong, W.~Guo, Y.~Chen, X.~Xing, Y.~Zhang, and G.~Wang, ``Towards the
  detection of inconsistencies in public security vulnerability reports,'' in
  \emph{28th {USENIX} Security Symposium ({USENIX} Security 19)}.\hskip 1em
  plus 0.5em minus 0.4em\relax Santa Clara, CA: {USENIX} Association, Aug.
  2019, pp. 869--885.

\bibitem{Sen@2019}
R.~Sen, J.~Choobineh, and S.~Kumar, ``Determinants of software vulnerability
  disclosure timing,'' \emph{Production and Operations Management}, vol.~29, 10
  2019.

\bibitem{Wang2020AnES}
X.~Wang, K.~Sun, A.~Batcheller, and S.~Jajodia, ``An empirical study of secret
  security patch in open source software,'' in \emph{Adaptive Autonomous Secure
  Cyber Systems}, 2020.

\bibitem{Wang@2020}
H.~Wang, G.~Ye, Z.~Tang, S.~H. Tan, S.~Huang, D.~Fang, Y.~Feng, L.~Bian, and
  Z.~Wang, ``Combining graph-based learning with automated data collection for
  code vulnerability detection,'' \emph{IEEE Transactions on Information
  Forensics and Security}, vol.~PP, 11 2020.

\bibitem{Zhou@2017}
Y.~Zhou and A.~Sharma, ``Automated identification of security issues from
  commit messages and bug reports,'' in \emph{Proceedings of the 2017 11th
  Joint Meeting on Foundations of Software Engineering}, ser. ESEC/FSE
  2017.\hskip 1em plus 0.5em minus 0.4em\relax Association for Computing
  Machinery, 2017, p. 914–919.

\bibitem{Sabetta@2018}
A.~Sabetta and M.~Bezzi, ``A practical approach to the automatic classification
  of security-relevant commits,'' 07 2018.

\bibitem{zolanvari2019machine}
M.~Zolanvari, M.~A. Teixeira, L.~Gupta, K.~M. Khan, and R.~Jain, ``Machine
  learning-based network vulnerability analysis of industrial internet of
  things,'' \emph{IEEE Internet of Things Journal}, vol.~6, no.~4, pp.
  6822--6834, 2019.

\bibitem{liu2019deepbalance}
S.~Liu, G.~Lin, Q.-L. Han, S.~Wen, J.~Zhang, and Y.~Xiang, ``Deepbalance:
  Deep-learning and fuzzy oversampling for vulnerability detection,''
  \emph{IEEE Transactions on Fuzzy Systems}, vol.~28, no.~7, pp. 1329--1343,
  2019.

\bibitem{bussone2015role}
A.~Bussone, S.~Stumpf, and D.~O'Sullivan, ``The role of explanations on trust
  and reliance in clinical decision support systems,'' in \emph{2015
  international conference on healthcare informatics}.\hskip 1em plus 0.5em
  minus 0.4em\relax IEEE, 2015, pp. 160--169.

\bibitem{nourani2020don}
M.~Nourani, C.~Roy, T.~Rahman, E.~D. Ragan, N.~Ruozzi, and V.~Gogate, ``Don't
  explain without verifying veracity: An evaluation of explainable ai with
  video activity recognition,'' \emph{arXiv preprint arXiv:2005.02335}, 2020.

\bibitem{Mohseni2018ASO}
S.~Mohseni, N.~Zarei, and E.~D. Ragan, ``A survey of evaluation methods and
  measures for interpretable machine learning,'' \emph{ArXiv}, vol.
  abs/1811.11839, 2018.

\bibitem{lee2015working}
M.~K. Lee, D.~Kusbit, E.~Metsky, and L.~Dabbish, ``Working with machines: The
  impact of algorithmic and data-driven management on human workers,'' in
  \emph{Proceedings of the 33rd annual ACM conference on human factors in
  computing systems}, 2015, pp. 1603--1612.

\bibitem{tang2012etrust}
J.~Tang, H.~Gao, H.~Liu, and A.~Das~Sarma, ``etrust: Understanding trust
  evolution in an online world,'' in \emph{Proceedings of the 18th ACM SIGKDD
  international conference on Knowledge discovery and data mining}, 2012, pp.
  253--261.

\bibitem{tintarev2011designing}
N.~Tintarev and J.~Masthoff, ``Designing and evaluating explanations for
  recommender systems,'' in \emph{Recommender systems handbook}.\hskip 1em plus
  0.5em minus 0.4em\relax Springer, 2011, pp. 479--510.

\bibitem{codeql}
CodeQL, \url{https://codeql.github.com/}, 2022, accessed: 2022-04-23.

\bibitem{9796256}
M.~Fu and C.~Tantithamthavorn, ``Linevul: A transformer-based line-level
  vulnerability prediction,'' in \emph{2022 IEEE/ACM 19th International
  Conference on Mining Software Repositories (MSR)}, 2022, pp. 608--620.

\bibitem{wattanakriengkrai2020predicting}
S.~Wattanakriengkrai, P.~Thongtanunam, C.~Tantithamthavorn, H.~Hata, and
  K.~Matsumoto, ``Predicting defective lines using a model-agnostic
  technique,'' \emph{IEEE Transactions on Software Engineering}, 2020.

\bibitem{pornprasit2021jitline}
C.~Pornprasit and C.~K. Tantithamthavorn, ``Jitline: A simpler, better, faster,
  finer-grained just-in-time defect prediction,'' in \emph{2021 IEEE/ACM 18th
  International Conference on Mining Software Repositories (MSR)}.\hskip 1em
  plus 0.5em minus 0.4em\relax IEEE, 2021, pp. 369--379.

\bibitem{ghassemi2021false}
M.~Ghassemi, L.~Oakden-Rayner, and A.~L. Beam, ``The false hope of current
  approaches to explainable artificial intelligence in health care,'' \emph{The
  Lancet Digital Health}, vol.~3, no.~11, pp. e745--e750, 2021.

\bibitem{linardatos2020explainable}
P.~Linardatos, V.~Papastefanopoulos, and S.~Kotsiantis, ``Explainable ai: A
  review of machine learning interpretability methods,'' \emph{Entropy},
  vol.~23, no.~1, p.~18, 2020.

\bibitem{zhang2020effect}
Y.~Zhang, Q.~V. Liao, and R.~K. Bellamy, ``Effect of confidence and explanation
  on accuracy and trust calibration in ai-assisted decision making,'' in
  \emph{Proceedings of the 2020 Conference on Fairness, Accountability, and
  Transparency}, 2020, pp. 295--305.

\bibitem{arora2022explain}
S.~Arora, D.~Pruthi, N.~Sadeh, W.~W. Cohen, Z.~C. Lipton, and G.~Neubig,
  ``Explain, edit, and understand: Rethinking user study design for evaluating
  model explanations,'' in \emph{Proceedings of the AAAI Conference on
  Artificial Intelligence}, vol.~36, no.~5, 2022, pp. 5277--5285.

\bibitem{wiegreffe2021reframing}
S.~Wiegreffe, J.~Hessel, S.~Swayamdipta, M.~Riedl, and Y.~Choi, ``Reframing
  human-ai collaboration for generating free-text explanations,'' \emph{arXiv
  preprint arXiv:2112.08674}, 2021.

\bibitem{jimenez2020drug}
J.~Jim{\'e}nez-Luna, F.~Grisoni, and G.~Schneider, ``Drug discovery with
  explainable artificial intelligence,'' \emph{Nature Machine Intelligence},
  vol.~2, no.~10, pp. 573--584, 2020.

\bibitem{alonso2020interactive}
J.~M. Alonso, S.~Barro, A.~Bugar{\'\i}n, K.~v. Deemter, C.~Gardent, A.~Gatt,
  E.~Reiter, C.~Sierra, M.~Theune, N.~Tintarev \emph{et~al.}, ``Interactive
  natural language technology for explainable artificial intelligence,'' in
  \emph{International Workshop on the Foundations of Trustworthy AI Integrating
  Learning, Optimization and Reasoning}.\hskip 1em plus 0.5em minus 0.4em\relax
  Springer, 2020, pp. 63--70.

\bibitem{CVEtemplate}
{CVE Request Template},
  \url{http://cveproject.github.io/docs/content/key-details-phrasing.pdf},
  2022, accessed: 2022-04-23.

\bibitem{sun2021generating}
J.~Sun, Z.~Xing, L.~Z. Sherry~Xu, and Q.~Lu, ``Heterogeneous vulnerability
  report traceability recovery by vulnerability aspect matching,'' in
  \emph{Proceedings of the ACM/IEEE 38th International Conference on Software
  Maintenance and Evolution}, ser. ICSME '22, 2022.

\bibitem{sun2021generating_old}
J.~Sun, Z.~Xing, H.~Guo, D.~Ye, X.~Li, X.~Xu, and L.~Zhu, ``Generating
  informative cve description from exploitdb posts by extractive
  summarization,'' \emph{arXiv preprint arXiv:2101.01431}, 2021.

\bibitem{feng2020codebert}
Z.~Feng, D.~Guo, D.~Tang, N.~Duan, X.~Feng, M.~Gong, L.~Shou, B.~Qin, T.~Liu,
  D.~Jiang \emph{et~al.}, ``Codebert: A pre-trained model for programming and
  natural languages,'' \emph{arXiv preprint arXiv:2002.08155}, 2020.

\bibitem{Tomcat}
Tomcat, \url{https://tomcat.apache.org/}, 2022, accessed: 2022-04-23.

\bibitem{Snyk}
{Snyk Databases}, \url{https://snyk.io/vuln}, 2022, accessed: 2022-04-23.

\bibitem{Liu@icse2020}
B.~Liu, G.~Meng, W.~Zou, Q.~Gong, F.~Li, M.~Lin, D.~Sun, W.~Huo, and C.~Zhang,
  ``A large-scale empirical study on vulnerability distribution within projects
  and the lessons learned,'' in \emph{Proceedings of the ACM/IEEE 42nd
  International Conference on Software Engineering}, ser. ICSE '20.\hskip 1em
  plus 0.5em minus 0.4em\relax New York, NY, USA: Association for Computing
  Machinery, 2020, p. 1547–1559.

\bibitem{cwe}
{Common Weakness Enumeration}, \url{https://cwe.mitre.org/}, 2022, accessed:
  2022-04-23.

\bibitem{IBM}
{IBM X-Force},
  \url{https://exchange.xforce.ibmcloud.com/activity/list?filter=Vulnerabilities},
  2022, accessed: 2022-04-23.

\bibitem{Exploitdb}
{Exploit Database}, \url{https://www.exploit-db.com/}, 2022, accessed:
  2022-04-23.

\bibitem{Openwall}
{Openwall mailing list archives}, \url{https://www.openwall.com/lists/}, 2022,
  accessed: 2022-04-23.

\bibitem{sak2014long}
H.~Sak, A.~W. Senior, and F.~Beaufays, ``Long short-term memory recurrent
  neural network architectures for large scale acoustic modeling,'' 2014.

\bibitem{greff2016lstm}
K.~Greff, R.~K. Srivastava, J.~Koutn{\'\i}k, B.~R. Steunebrink, and
  J.~Schmidhuber, ``Lstm: A search space odyssey,'' \emph{IEEE transactions on
  neural networks and learning systems}, vol.~28, no.~10, pp. 2222--2232, 2016.

\bibitem{zhou2016deep}
J.~Zhou, Y.~Cao, X.~Wang, P.~Li, and W.~Xu, ``Deep recurrent models with
  fast-forward connections for neural machine translation,'' \emph{Transactions
  of the Association for Computational Linguistics}, vol.~4, pp. 371--383,
  2016.

\bibitem{graves2013hybrid}
A.~Graves, N.~Jaitly, and A.-r. Mohamed, ``Hybrid speech recognition with deep
  bidirectional lstm,'' in \emph{2013 IEEE workshop on automatic speech
  recognition and understanding}.\hskip 1em plus 0.5em minus 0.4em\relax IEEE,
  2013, pp. 273--278.

\bibitem{zhou2021spi}
Y.~Zhou, J.~K. Siow, C.~Wang, S.~Liu, and Y.~Liu, ``Spi: Automated
  identification of security patches via commits,'' \emph{arXiv preprint
  arXiv:2105.14565}, 2021.

\bibitem{Zhen@2020}
Z.~Li, D.~Zou, S.~Xu, Z.~Chen, Y.~Zhu, and H.~Jin, ``Vuldeelocator: A deep
  learning-based fine-grained vulnerability detector,'' 01 2020.

\bibitem{transformer}
A.~Vaswani, N.~Shazeer, N.~Parmar, J.~Uszkoreit, L.~Jones, A.~N. Gomez,
  L.~Kaiser, and I.~Polosukhin, ``Attention is all you need,'' in
  \emph{Proceedings of the 31st International Conference on Neural Information
  Processing Systems}, ser. NIPS'17, 2017, p. 6000–6010.

\bibitem{roziere2020unsupervised}
B.~Roziere, M.-A. Lachaux, L.~Chanussot, and G.~Lample, ``Unsupervised
  translation of programming languages,'' \emph{Advances in Neural Information
  Processing Systems}, vol.~33, pp. 20\,601--20\,611, 2020.

\bibitem{NEURIPS2021}
M.-A. Lachaux, B.~Roziere, M.~Szafraniec, and G.~Lample, ``Dobf: A
  deobfuscation pre-training objective for programming languages,'' in
  \emph{Advances in Neural Information Processing Systems}, vol.~34.\hskip 1em
  plus 0.5em minus 0.4em\relax Curran Associates, Inc., 2021, pp.
  14\,967--14\,979.

\bibitem{bridle1990probabilistic}
J.~S. Bridle, ``Probabilistic interpretation of feedforward classification
  network outputs, with relationships to statistical pattern recognition,'' in
  \emph{Neurocomputing}.\hskip 1em plus 0.5em minus 0.4em\relax Springer, 1990,
  pp. 227--236.

\bibitem{mikolov2013distributed}
T.~Mikolov, I.~Sutskever, K.~Chen, G.~S. Corrado, and J.~Dean, ``Distributed
  representations of words and phrases and their compositionality,''
  \emph{Advances in neural information processing systems}, vol.~26, 2013.

\bibitem{he2016deep}
K.~He, X.~Zhang, S.~Ren, and J.~Sun, ``Deep residual learning for image
  recognition,'' in \emph{Proceedings of the IEEE conference on computer vision
  and pattern recognition}, 2016, pp. 770--778.

\bibitem{CVEDetail}
{CVE Details}, \url{https://www.cvedetails.com/}, 2023, accessed: 2023-01-18.

\bibitem{lin2004rouge}
C.-Y. Lin, ``Rouge: A package for automatic evaluation of summaries,'' in
  \emph{Text summarization branches out}, 2004, pp. 74--81.

\bibitem{nourani2021anchoring}
M.~Nourani, C.~Roy, J.~E. Block, D.~R. Honeycutt, T.~Rahman, E.~Ragan, and
  V.~Gogate, ``Anchoring bias affects mental model formation and user reliance
  in explainable ai systems,'' in \emph{26th International Conference on
  Intelligent User Interfaces}, 2021, pp. 340--350.

\bibitem{Papagiannis@2011}
I.~Papagiannis, M.~Migliavacca, and P.~Pietzuch, ``Php aspis: Using partial
  taint tracking to protect against injection attacks,'' 06 2011, pp. 2--2.

\bibitem{Viega@2000}
J.~Viega, J.~Bloch, Y.~Kohno, and G.~McGraw, ``Its4: a static vulnerability
  scanner for c and c++ code,'' in \emph{Proceedings 16th Annual Computer
  Security Applications Conference (ACSAC'00)}, 2000, pp. 257--267.

\bibitem{Younan04codeinjection}
Y.~Younan, W.~Joosen, and F.~Piessens, ``Code injection in c and c++ : A survey
  of vulnerabilities and countermeasures,'' DEPARTEMENT COMPUTERWETENSCHAPPEN,
  KATHOLIEKE UNIVERSITEIT LEUVEN, Tech. Rep., 2004.

\bibitem{Chen@2020}
Y.~Chen, A.~E. Santosa, A.~M. Yi, A.~Sharma, A.~Sharma, and D.~Lo, ``A machine
  learning approach for vulnerability curation,'' in \emph{Proceedings of the
  17th International Conference on Mining Software Repositories}, ser. MSR
  '20.\hskip 1em plus 0.5em minus 0.4em\relax Association for Computing
  Machinery, 2020, p. 32–42.

\bibitem{Ramsauer@2020}
R.~Ramsauer, L.~Bulwahn, D.~Lohmann, and W.~Mauerer, ``The sound of silence:
  Mining security vulnerabilities from secret integration channels in
  open-source projects,'' in \emph{Proceedings of the 2020 ACM SIGSAC
  Conference on Cloud Computing Security Workshop}, ser. CCSW'20, 2020, p.
  147–157.

\bibitem{weld2019challenge}
D.~S. Weld and G.~Bansal, ``The challenge of crafting intelligible
  intelligence,'' \emph{Communications of the ACM}, vol.~62, no.~6, pp. 70--79,
  2019.

\bibitem{simonyan2013deep}
K.~Simonyan, A.~Vedaldi, and A.~Zisserman, ``Deep inside convolutional
  networks: Visualising image classification models and saliency maps,''
  \emph{arXiv preprint arXiv:1312.6034}, 2013.

\bibitem{zeiler2014visualizing}
M.~D. Zeiler and R.~Fergus, ``Visualizing and understanding convolutional
  networks,'' in \emph{European conference on computer vision}.\hskip 1em plus
  0.5em minus 0.4em\relax Springer, 2014, pp. 818--833.

\bibitem{berkovsky2017recommend}
S.~Berkovsky, R.~Taib, and D.~Conway, ``How to recommend? user trust factors in
  movie recommender systems,'' in \emph{Proceedings of the 22nd international
  conference on intelligent user interfaces}, 2017, pp. 287--300.

\bibitem{lakkaraju2016interpretable}
H.~Lakkaraju, S.~H. Bach, and J.~Leskovec, ``Interpretable decision sets: A
  joint framework for description and prediction,'' in \emph{Proceedings of the
  22nd ACM SIGKDD international conference on knowledge discovery and data
  mining}, 2016, pp. 1675--1684.

\bibitem{hohman2018visual}
F.~Hohman, M.~Kahng, R.~Pienta, and D.~H. Chau, ``Visual analytics in deep
  learning: An interrogative survey for the next frontiers,'' \emph{IEEE
  transactions on visualization and computer graphics}, vol.~25, no.~8, pp.
  2674--2693, 2018.

\bibitem{tantithamthavorn2021explainable}
C.~K. Tantithamthavorn and J.~Jiarpakdee, ``Explainable ai for software
  engineering,'' in \emph{2021 36th IEEE/ACM International Conference on
  Automated Software Engineering (ASE)}.\hskip 1em plus 0.5em minus 0.4em\relax
  IEEE, 2021, pp. 1--2.

\bibitem{jiarpakdee2020empirical}
J.~Jiarpakdee, C.~K. Tantithamthavorn, H.~K. Dam, and J.~Grundy, ``An empirical
  study of model-agnostic techniques for defect prediction models,'' \emph{IEEE
  Transactions on Software Engineering}, vol.~48, no.~1, pp. 166--185, 2020.

\bibitem{li2021vulnerability}
Y.~Li, S.~Wang, and T.~N. Nguyen, ``Vulnerability detection with fine-grained
  interpretations,'' in \emph{Proceedings of the 29th ACM Joint Meeting on
  European Software Engineering Conference and Symposium on the Foundations of
  Software Engineering}, 2021, pp. 292--303.

\bibitem{jiarpakdee2021practitioners}
J.~Jiarpakdee, C.~K. Tantithamthavorn, and J.~Grundy, ``Practitioners’
  perceptions of the goals and visual explanations of defect prediction
  models,'' in \emph{2021 IEEE/ACM 18th International Conference on Mining
  Software Repositories (MSR)}.\hskip 1em plus 0.5em minus 0.4em\relax IEEE,
  2021, pp. 432--443.

\bibitem{huang2021cosqa}
J.~Huang, D.~Tang, L.~Shou, M.~Gong, K.~Xu, D.~Jiang, M.~Zhou, and N.~Duan,
  ``Cosqa: 20,000+ web queries for code search and question answering,''
  \emph{arXiv preprint arXiv:2105.13239}, 2021.

\bibitem{nguyen2021regvd}
V.-A. Nguyen, D.~Q. Nguyen, V.~Nguyen, T.~Le, Q.~H. Tran, and D.~Phung,
  ``Regvd: Revisiting graph neural networks for vulnerability detection,''
  \emph{arXiv preprint arXiv:2110.07317}, 2021.

\bibitem{shi2022enhancing}
E.~Shi, W.~Gub, Y.~Wang, L.~Du, H.~Zhang, S.~Han, D.~Zhang, and H.~Sun,
  ``Enhancing semantic code search with multimodal contrastive learning and
  soft data augmentation,'' \emph{arXiv preprint arXiv:2204.03293}, 2022.

\bibitem{goel2022cross}
D.~Goel, R.~Grover, and F.~H. Fard, ``On the cross-modal transfer from natural
  language to code through adapter modules,'' \emph{arXiv preprint
  arXiv:2204.08653}, 2022.

\end{thebibliography}
\end{document}